\documentclass{ar-1col}
\usepackage{xspace}
\usepackage{amssymb,amsmath}
\usepackage[authoryear,round]{natbib}
\usepackage{biblio}
\usepackage[total={30.5pc,47pc},centering]{geometry}
\jname{Annu. Rev. Astron. Astrophys.}
\jvol{XX}
\jyear{2018}

\begin{document}
\markboth{Pedro G. Ferreira}{Cosmological Gravity}

\title{Cosmological Tests of Gravity}

\author{Pedro G. Ferreira$^1$ 
\affil{$^1$Astrophysics, University of Oxford,  Oxford OX1 3RH, UK; email: pedro.ferreira@physics.ox.ac.uk}
}

\begin{abstract}
Cosmological observations are beginning to reach a level of precision that allows us to test some of the most fundamental assumptions in our working model of the Universe. One such as assumption is that gravity is governed by the General Theory of Relativity. In this review we discuss how one might go about extending General Relativity and how such extensions can be described in a unified way  on large scales. This allows us to describe the phenomenology of modified gravity in the growth and morphology of the large scale structure of the Universe. On smaller scales we explore the physics of gravitational screening and how it might manifest itself in galaxies, clusters and, more generally, in the cosmic web. We then analyse the current constraints from large scale structure and conclude by discussing the future prospects of the field in light of the plethora of surveys currently being planned. 

The key results are:
\begin{itemize}
\item there are a plethora of alternative theories of gravity which are\\  restricted by fundamental physics considerations; 
\item there is now a well established formalism for describing\\ cosmological perturbations in the linear regime for general\\ theories of gravity;
\item gravitational screening can mask modifications to general relativity\\ on small scales but may, itself, lead to distinctive signatures in the\\ large scale structure of the Universe;
\item current constraints on both linear and non-linear scales may be\\ affected by systematic uncertainties which limit our ability to rule\\ out alternatives to General Relativity;
\item the next generation of cosmological surveys will dramatically improve\\ constraints on General Relativity, by up to two orders of magnitude.
\end{itemize}
\end{abstract}

\begin{keywords}
General Relativity, Cosmology, Large Scale Structure, Cosmic Microwave Background, Early Universe
\end{keywords}
\maketitle

\tableofcontents

\section{Why test gravity?}
\label{introduction}
\subsection{The success and the uniqueness of General Relativity.}
The success of the standard model of fundamental physics is unquestionable and unassailable. There is a complete model for the strong and electro-weak forces (dubbed the standard model -- SM) which has passed every single test to which it has been subjected. While some minor anomalies have cropped up, they are not sufficiently significant to overhaul the main picture: a gauge theory of interactions between fermions where electro-weak symmetry is spontaneously broken by a scalar field. The remaining force, gravity, is perfectly described by the general theory of relativity (GR), a theory of a dynamical space time where general covariance  (also called general coordinate invariance or diffeomorphism invariance) plays a fundamental role. Again, observations repeatedly confirm that GR works  in a number of regimes. Observations and experiments on the scale of the Solar System using lunar laser ranging, satellite missions such as Cassini or exquisite measurements of planetary orbits compete with measurements of individual and binary millisecond pulsars as well as triple systems to obtain exquisite constraints of gravity \citep{Will2014}.
With the advent of gravitational wave astronomy and black hole imaging, it will be possible to, for the first time, find constraints in the strong gravity regime.

When extrapolated to the largest, cosmological, scales, the SM and GR seem to accurately describe the history and evolution of the universe, as well as the large scale structure of space-time and matter. In that case, the SM and GR have been extrapolated to length scales and energy scales which are well beyond any current experimentally (or astrophysically) acessible regime. In other words,  two theories which pass tests in the laboratory and in the Solar System with flying colours are used to predict what is observed on length scales which are up to fifteen orders of magnitude greater. And the predictions match observations with remarkable consistency: the expansion rate of the universe, the morphology of large scale structure, the abundance of different types of particles, can all be consistently modelled with the current cosmological model. Furthermore, it allows us to predict that there is a dark sector to the Universe, i.e. that in the context of SM and GR, we should expect $95\%$ of the energy density of the Universe to consist of some exotic form of material -- dark matter and dark energy. The simplest formulation of this cosmological model has been called $\Lambda$- Cold Dark Matter ($\Lambda$CDM) where $\Lambda$ represents the cosmological constant. It is now customary to express cosmological results in terms of constraints on $\Lambda$CDM, some of which can reach sub-percent accuracy.

If one delves into the structure of gravity, there are compelling reasons to believe that GR is the unique theory that can explain it. Assuming that gravity is mediated by a spin-2 carrier -- a transverse, traceless field -- there are a number of arguments that show that GR is the non-linear theory that can correctly explain its dynamics. In a series of lectures and papers \citep{Feynman1995,Weinberg1964,Deser1970} it was shown that, beginning with the massless, Fierz-Pauli action for a spin-2 field, $h_{\alpha\beta}$ \citep{FP1939}, the requirement of general covariance  implies that one needs to consider non-linear corrections and that, when all of them are correctly taken into account, means that the fundamental action is unique and is that of GR.
An alternative, and more classical, viewpoint \citep{Lovelock1972} shows that the only second-order, local gravitational field equations derivable from an action containing solely the 4D metric tensor (plus related tensors) are the
Einstein field equations with a cosmological constant. This is backed up by a similar conclusion from the geometrodynamic point of view \citep{Hojman1976}.

\subsection{Hints from the late and early Universe.}
While the evidence for the SM and GR as an integral component of the current working model of the Universe is extremely strong, it is important to step back and ponder the way forward. Indeed, with these core theories we can make remarkable predictions on cosmological scales and, in particular, it leads to extremely strong evidence for a dark sector. The evidence is so compelling that the quest to understand this dark sector is the scientific driver of a number of large scale observational programmes. Nevertheless, given that there is no other (non-gravitational) evidence for the dark sector, it is a matter of common sense to question some of the fundamental assumptions that go into the evidence. And the main assumption is that GR is the underlying theory of gravity. An alternative point of view then, is that evidence for the dark sector may signal a break down of GR on cosmological scales, at late time. 

There is tantalizing evidence for deviation from general relativity if one attempts to place constraints on the physics of the very early universe. From observations of the anisotropy of the CMB, it is now possible to place extremely stringent constraints on the statistics of primordial fluctuations. The simplest assumption is that these fluctuations can be characterized in terms of an overall amplitude for both scalar and tensor components along with spectral indices for their spatial morphology. Current constraints on the spectral index place it close to, but not exactly at, scale invariance; they also place an upper-bound on the amplitude of tensor fluctuations (in the form of primordial gravitational waves) relative to scalar fluctuations (in the form of density perturbations) \citep{Akrami2018a}. The most favoured theory that can explain these observations is inflation: a period of accelerated expansion in the early universe, amplifying microscopic quantum fluctuations into macroscopic cosmological perturbations.  Constraints from the CMB greatly restrict the range of models which are observationally viable. Remarkably, those that are favoured involve some modification to GR at very early times, either through a non-minimal coupling between the inflaton scalar field and the Ricci tensor or through the inclusion of higher order powers of the Ricci tensor in the fundamental action \citep{Kaiser2014}.

\subsection{The (almost) strong field regime.}
To some extent it is unsurprising that modifications should arise in a regime of high curvature. Purely from an effective field theory point of view \citep{Donoghue1994,Burgess2003}, one should expect corrections to GR to emerge in this regime -- the Einstein-Hilbert action should have corrections proportional to $R^2$, $R^{\mu\nu}R_{\mu\nu}$ and so one. Furthermore, in such high curvature regimes, any extra fields (or degrees of freedom) which are frozen at low energies will, most likely, be dynamical and contribute to any non-standard gravitational dynamics. It is then natural to look for modifications away from the weak field regime: while it is practically impossible to access high curvature regimes (apart from the early Universe or in the cores of black holes) one can be slightly less ambitious and look at the regime where the gravitational potential, $\Phi$ is of order unity. And, as luck would have it, two brand new windows on gravitational physics have opened up in this regime: gravitational wave detection and black hole imaging.

The advanced Laser Interferometric Gravitational Observatory (aLIGO) has observed a number of coalescing compact objects through their gravitational wave emission \citep{Abbott2016a}. A number of merging black hole and neutron star binaries have been used to place stringent constraints on gravitational physics -- thus far these constraints have focused on internal consistency tests in the waveform models where the signal is dominated by the weak field regime during the inspiral and after the merger \citep{Abbott2016b}. But as the number of events build up, and analytic understanding of the merger improves, it should be possible to dig into the merger where the gravitational potential becomes appreciable. It should also be possible to extract information from the ring-down of the merger where signatures of deviation from GR may become evident in the quasi-normal mode spectrum. From this spectrum it will be possible, for example, to test the no-hair theorem or probe for the presence of extra degrees of freedom \citep{Berti2015}. Alternatively, by studying mergers which are associated with electromagnetic counterparts, it is possible to place constraints on the propagation of gravitational waves and place constraints on the gravitational sector. Indeed, that has already been done to great effect \citep{Baker2017,Creminelli2017,Ezquiaga2017} with the recently detected binary neutron star merger GW170817 \citep{Abbott2017}.

A coordinated effort by the consortia of telescopes that make the Event Horizon Telescope to observe Sgr A$^*$ at the center of the Galaxy, will lead to the first image of a black hole event horizon (https://eventhorizontelescope.org) . By analyzing the structure of the accretion flow and how it wraps around the dark inner shadow, it will be possible to probe the structure of space-time close to the Schwarzschild radius and test whether it is consistent with GR, i.e. whether it is has a Kerr geometry. Preliminary measurements have already picked up structure on the scale of the event horizon \citep{Doeleman2008} and, while one should expect the intricacies of the physics of accretion flows to complicate the analysis, it will be possible, for the first time, to look for evidence for deviations to GR close to (but not in) the strong gravity regime \citep{Johannsen2010,Broderick2013}.

\subsection{Modified gravity in cosmology: a roadmap.}
Let us now turn our sights onto cosmological scales at late times. A first guess would be that deviations from GR on these scales would be unexpected. While the expanding Universe is, in some sense, a core result of GR or any relativistic form of gravity, and can be thought of as in the strong field regime, any fluctuations away from the homogeneous and isotropic space-time are extremely weak. Naively one would expect that it would be very difficult to construct modifications to GR in the regime where the curvature is small -- in fact one might even think it problematic given that Minkowski space is the limit of zero curvature and we believe it to be an excellent approximation in a number of classical and quantum scenarios.

The past few decades have shown that naive expectations about the cosmological regime don't necessarily bear out. For a start, there has been an explosion of theoretical ideas leading to models that modify gravity in the infra-red while leaving it consistent with GR on astrophysical scales \citep{Clifton2012}. These have involved exploring the various loop-holes to Lovelock's theorem -- higher dimensions and derivatives, extra fields and non-locality. The various proposals can all be rephrased in terms of extra fields which couple non-minimally to gravity and lead to new fifth forces.  Non-linear, classical effects can lead to screening of gravitational fifth forces in regimes of high gravitational potential and curvature while leaving them unscreened on large scales and in regimes of low curvature \citep{Vainshtein1972,Khoury2004,Hinterbichler2010}. Constraints on fifth forces are remarkable \citep{Adelberger2003} but are restricted to small scales (compared to cosmology). Hence, looking for modifications to general relativity on large scales is tantamount to looking for the effects of fifth forces in an unscreened regime. This is uncharted territory and one of the few pristine arenas where we might find evidence for new physics.

The search for dark energy has also become a search for modifications to general relativity on large scales. All the main observational programmes currently being planned have, as one part of their science cases, the goal of testing gravity using a range of different observables with a variety of tracers of large scale structure. The purpose of this review is to establish the core ideas underpinning such a programme. I will describe the basic ideas and tools that go into extracting meaningful constraints from cosmological data.

The structure of this review is as follows. In Section \ref{cosmology} I briefly review current cosmology, describing the key aspects of gravitational physics that play a role in the evolution of the Universe. In Section \ref{modgrav} I discuss how gravity can be modified on cosmological scales, with a particular emphasis on how this involves extra degrees of freedom and fields. In Section \ref{screening} I work through the basics of gravitational screening and how it might mask modifications to gravity in certain regimes. In Section \ref{LSS} I talk through the various, existing constraints on gravity with large scale observables using existing surveys while in Section \ref{alternatives} I focus on constraints involving the non-linear regime and environmental effects. Finally, in Section \ref{future}, I map out the future of the field.

\section{Modern Cosmology.}
\label{cosmology}
\subsection{The expanding Universe.}
 
The evolution of the Universe can be predicted from the field equations of GR, under the assumption that the metric of space-time, $g_{\alpha\beta}$ is homogeneous and isotropic, $g_{\alpha\beta}={\rm diag}(-1, a^2\delta_{ij})$ where $a$ is the scale factor and only depends on time (for simplicity we will assume spatial flatness throughout). The dynamics of $a$ is given by
\begin{eqnarray}
H^2\equiv\left(\frac{\dot a}{a}\right)^2=\frac{8\pi G}{3}{\bar \rho} \label{FRW1}
\end{eqnarray}
where overdot denotes a derivative with regards to time and ${\bar \rho}$ is the mean density of the Universe. The acceleration of the scale factor is given by
\begin{eqnarray}
\frac{\ddot a}{a}=-\frac{4\pi G}{3}({\bar \rho}+3{\bar P}) \label{FRW2}
\end{eqnarray}
where ${\bar P}$ is the total mean pressure of the Universe. Equation \ref{FRW1} can be rewritten in terms of the fractional energy densities of the various components, $\Omega_X={8\pi G \bar \rho}_X/(3H_0^2)$ (where $X$ can be: radiation, R, baryons, B, cold dark matter, C, dark energy, DE) such that
\begin{eqnarray}
H^2(a)=H^2_0\left[\frac{\Omega_{\rm R}}{a^4}+\frac{\Omega_{\rm B}+\Omega_{\rm C}}{a^3}+\frac{\Omega_{\rm DE}}{a^{3(w+1)}}\right]
\end{eqnarray}
(where for simplicity we have assumed here a constant $w$). Clearly, a measurement of $H$ as a function of time or scale factor, $a$, can be used to place constraints on the $\Omega_X$ and on the equation of state of dark energy, $w={\bar P}_{\rm DE}/{\bar \rho}_{\rm DE}$.

Distant objects will have their light redshifted: $\nu_{\rm emitted}/\nu_{\rm observed}=a_0/a\equiv 1+z$ where the ratio of the emitted and observed frequencies depend on the scale factor at emission (a) and at the observer $a_0$, and where we have defined the redshift $z$. The relation between distance and redshift can be reconstructed through Equation \ref{FRW1}. For example, for a Euclidean universe, the angular-diameter distance, $D_A$, (i.e. the distance reconstructed from knowing the physical size and measuring the angle it subtends in the sky) is given by
\begin{eqnarray}
D_A(z)=\frac{1}{1+z}\int_0^z\frac {dz}{H(z)}  
\end{eqnarray} 
and the corresponding luminosity distance is given by $D_L=(1+z)^2D_A$. By measuring redshifts and distances of stars, galaxies, supernovae and matching them to $D_A(z)$ or $D_L(z)$ at $z$ close to $0$ we can estimate $H(z=0)\equiv H_0$; by measuring more distant objects, we can infer other cosmological parameters such as the $\Omega_X$'s and $w$.

The expansion of the Universe has an impact on matter and radiation. Throughout its history, the Universe has undergone a series of transitions as the overall temperature decreases. One of the most striking transitions occurs when the temperature is of a few thousand degrees Kelvin, corresponding to a few tenths of an electron-Volt. At that time, when the Universe was about 380,000 years old, at a redshift, $z\simeq 1100$, free electrons and protons combined to form neutral Hydrogen. As a result, the ambient medium which was, until then, optically thick, became transparent; photons which last scattered with residual electrons and protons, remained free to propagate along null geodesics. As a result, we are constantly bombarded by a relic bath of radiation with black-body distribution with temperature, ${\bar T}=2.735^o$ Kelvin and an almost perfectly isotropic distribution. Indeed, it is useful to think of these relic photons as having been emitted  at  time $t_*$ from the surface of sphere -- the surface of last scatter -- centred on us.

\subsection{Linear cosmological perturbation theory.}
It is, of course, with the study of large scale structure that cosmology has really taken off. Going beyond a homogeneous and isotropic universe, one needs to include inhomogeneities. This can be done by perturbing the metric, $g_{\alpha\beta}={\rm diag}[-(1+2\Phi), a^2(1-2\Psi)\delta_{ij})]$ (in the Newtonian gauge), the various densities, $\rho_X={\bar \rho}_X(1+\delta_X)$ and pressures, $P_X={\bar P}_X(1+\delta_{PX})$ and the temperature of the CMB, $T={\bar T}(1+\frac{\delta T}{\bar T})$; note that $\Phi$, $\Psi$, $\delta_X$ and $\delta_{PX}$ are functions of time, $t$ and position, ${\vec x}$ while $\frac{\delta T}{\bar T}$ is  function of $t$, ${\vec x}$ and direction ${\hat n}$.

The evolution of the perturbations can be found through cosmological perturbation theory: taking the full equations for gravity (the Einstein field equations) along with the equations for conservation of mass and momentum (the relativistic Euler equation) and perturbing to linear order. So, for example, in Fourier space, a combination of the $00$ and $0i$ component of the Einstein field equations yields something akin to the relativistic Newton-Poisson equation,
\begin{eqnarray}
-k^2\Phi=4\pi G \Delta\rho \label{NP}
\end{eqnarray}
with $\Delta\rho=\sum_X{\bar \rho}_X[\delta_X+3(1+w_X)\frac{H}{a}\frac{\theta_X}{k^2}]$ and $\theta_X={\vec \nabla}\cdot {\vec v}_X$ where ${\vec v}_X$ is the peculiar velocity of component $X$ and $k$ labels the Fourier mode. And the traceless, transverse component (in the absence of anisotropic stress) of the Einstein field equation give
\begin{eqnarray}
k^2(\Phi-\Psi)=0 \label{Slip}
\end{eqnarray}
The perturbed conservation equations (i.e. the linearized $\nabla_\mu T^{\mu\nu}=0$ equation, where $\nabla$ is the covariant derivative and $T^{\mu\nu}$ is the energy momentum tensor) can be used find additional evolution equations for $\delta_X$ and $\theta_X$. With a suitable definition of the  sound speed, $c^2_X$ one can replace $\delta_{PX}$.

\subsection{The observables: power spectra, growth rate and cosmic shear.}
When studying the perturbations around a homogeneous and isotropic background, statistics is paramount. First of all, and given that the (ensemble) average (or mean) of all linear perturbations is zero, variance plays a key role. Thus the relevant quantities that need to be considered are the power spectrum of density fluctuations, $P(k)$, and the angular power spectrum of CMB fluctuations, $C_\ell$, given by
\begin{eqnarray}
\langle \delta^*_{M}({\bf k}')\delta_{M}({\bf k})\rangle&=&(2\pi)^3P(k)\delta^{(3)}({\bf k}-{\bf k}') \nonumber \\
\langle a^{*}_{\ell' m'}a_{\ell m}\rangle&=&C_\ell\delta_{\ell\ell'}\delta_{mm'}
\end{eqnarray}
where $\delta_{M}({\bf k})$ is the Fourier transform of $\delta_B+\delta_C$, and $a_{\ell m}$ is the spherical harmonic transform of $T({\bf n})$ (or of any two-dimensional observable on a sphere). The various power spectra (and cross-spectra, i.e. the cross  correlation between different fields) are at the heart of the bulk of cosmological analysis.

\begin{figure*}[h]
\includegraphics[width=0.7\textwidth]{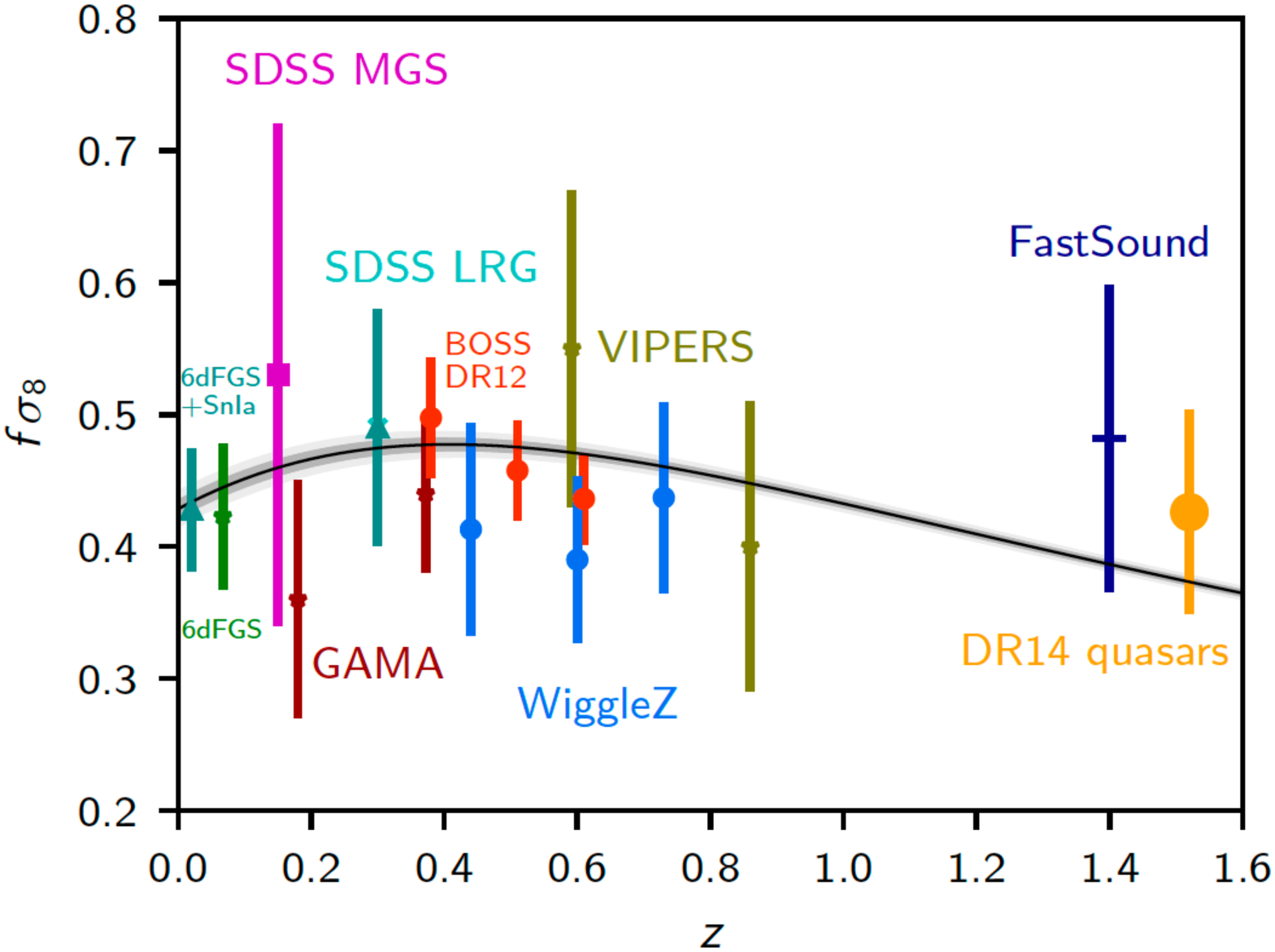}
\caption{The density weighted growth weight as a function of redshift \citep{Aghanim2018}. The shaded line is the extrapolated value for $\Lambda$CDM, marginalized over the parameters subjected to the Planck 2018 data. The shapes with error bars are a selection of independent measurements from large scale structure surveys.}
\label{growth}
\end{figure*}

Of particular importance is the rate of growth of cosmic structure. The linear density contrast can be factorized, $\delta_M({\bf k},t)=D(t)\delta({\bf k},t_0)$ where $t_0$ is the physical time today. In the case of a matter dominated Universe (or a Universe with a cosmological constant), the fact that the time dependence of the evolution is scale-independent is strictly satisfied. One can then define two useful quantities, the {\it growth rate},
\begin{eqnarray}
f\equiv\frac{d\ln D}{d\ln a}
\end{eqnarray}
and the "density weighted growth rate", $f\sigma_8$, where $\sigma^2_8=\langle(\delta M)^2/{\bar M}^2\rangle|_{8 h^{-1} {\rm Mpc}}$, i.e. the mass-variance on $8 h^{-1} {\rm Mpc}$ scales. These quantities map out the time dependence of gravitational collapse and will play a key role in cosmological tests of gravity. A fair representation of the current observational status of the growth can be found in Figure \ref{growth}: the shaded line shows the expected growth rate from CMB constraints (dominated by density and potential fluctuations at $z\simeq1100$) and projected forward assuming $\Lambda$CDM, which should be compared with individual and independent constraints (i.e. the various dots with errors bars) from galaxy surveys which are (necessarily) at late redshift. 

\begin{figure*}[h]
\includegraphics[width=\textwidth]{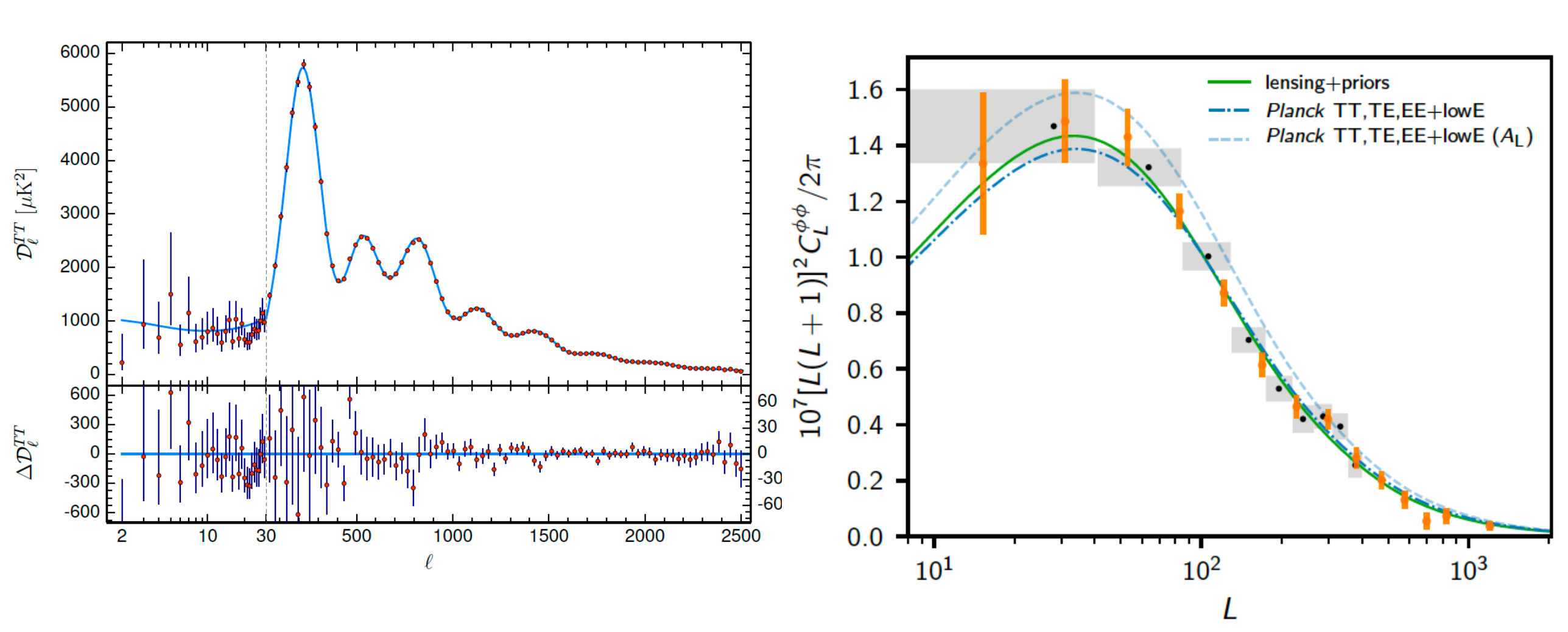}
\caption{Left hand panel: (top) the angular power spectrum of anisotropies of the CMB from the Planck 2018 data release \citep{Aghanim2018} and (bottom) the difference between the data and the best fit theoretical model ($\Lambda$CDM). Right hand panel: the angular power spectrum of the CMB lensing potential (orange) from Planck and a selection of best fit models to different combinations of additional data sets.}
\label{CMB}
\end{figure*}

The anisotropies in the CMB contain information about the radiation density contrast, $\delta_{\rm \gamma}$, the baryon velocity, ${\vec v}_{\rm B}$ and potential fluctuations, $\Phi$ at last scattering (known as the Sachs-Wolfe effect) as well as potential fluctuations from last scattering until now (known as the Integrated Sachs-Wolfe effect) through:
\begin{eqnarray}
\frac{\delta T}{T}({\hat n})=\frac{1}{3}\delta_\gamma|_*-{\vec v}_{\rm B}\cdot{\hat n}|_*+\Phi|_*-\int_{t_*}^{t_0}({\dot \Phi}+{\dot \Psi})\frac{dt}{a} \label{CMBeq}
\end{eqnarray}
where subscript $*$ means evaluated at the time of last scattering, $t_*$.
 Furthermore, the CMB will be slightly polarized due to the finite thickness of the surface of last scattering and by the reionized electrons at later times. In the left hand panel of Figure \ref{CMB} we show the angular power spectra of the anisotropy in CMB intensity and polarization; one can see distinctive features such as a series of oscillations due to the sound waves in the baryon-photon plasma before recombination as well as the damping on small scales (i.e. large $\ell$) that arises from the thickness of the surface of last scattering. 

The CMB will also be affected by the intervening structure, from the surface of last scatter until now. As the photons propagate through the perturbed space-time, they will be deflected and the structure of hot and cold spots will be distorted and deformed -- the CMB will be {\it lensed}. One way the effect will manifest itself is be distorting the angular power spectrum of the CMB; specifically it will very slightly smooth out the peaks and troughs arising from the baryon-photon oscillations generated before recombination. Another way lensing affects the CMB is by modifying the statistics of $\frac{\delta T}{T}$; what was originally a multivariate Gaussian will become non-Gaussian due to lensing and by measuring the 4-point function (or equivalently the trispectrum) and seeing how it deviates from the Gaussian value, it is possible to reconstruct the angular power spectrum of the lensing potential, $C^{\phi\phi}_\ell$.
In the right hand panel of Figure \ref{CMB} we show the angular power spectrum of the reconstructed lensing potential, i.e. of $C^{\phi\phi}_\ell$.

Distant galaxies will also be lensed by intervening structure. Measurements of distortions of the shapes of galaxies, i.e. changes in the projected ellipticities, and fluctuations in magnitudes, can be directly related to the distortion tensor
\begin{eqnarray}
\left( \begin{array}{cc}
1-\kappa-\gamma_1&-\gamma_2\\
-\gamma_2 &1-\kappa+\gamma_1  \end{array} \right)= \delta_{ij}+\int_0^{\chi}d\chi'(\Phi+\Psi)_{ij}[{\vec x}(\chi')]\chi'\left(1-\frac{\chi'}{\chi}\right)\label{lensing}
\end{eqnarray}
where the $\gamma_i$ are the shears and $\kappa$ is the convergence, $\chi=\int_{t_0}^t\frac{dt'}{a(t')}$ is the conformal distance and ($i$,$j$) are directions perpendicular to the direction of the light ray. One can attempt to reconstruct the gravitational potential probed by the incoming light rays by inverting equation \ref{lensing} and, for example, relating it to underlying mass distribution through the Poisson equation. This method has been successfully used for reconstructing the mass distribution of clusters of galaxies as well as for mapping out the large scale distribution of dark matter. Alternatively, one can study the auto and cross correlations of $\kappa$ and $\gamma_i$ as well as cross correlations with other measures of the galaxy density fluctuations.

\subsection{Small scales and non-gravitational physics.}
While much of what has been discussed is primarily applicable to the largest scales, where linear perturbation theory is applicable, the observables can also be used on smaller scales. Furthermore, we expect the smallest scales to have a larger statistical weight -- there are more modes contributing to any particular quantity. But while the statistical weight will improve, the underlying physics becomes more complex as non-linear gravitational collapse plays a role, allied with non gravitational physics. Baryons, through the role of gas and stars or highly energetic processes like supernovae and active galactic nuclei can play a significant role in moulding structure on scales of Megaparsecs and below \citep{Chisari2018}. 

There has been considerable progress in developing both analytic techniques (through different versions of perturbation theory \citep{Bernardeau2002} and effective field theory \citep{Carrasco2012}) and numerical algorithms (with N-body simulations, combined N-body and hydrodynamic simulations and, more recently, machine learning algorithms). In this, more complex regime, one can look at a more varied set of statistics, above and beyond the variances described above. For example, three point and four point statistics (i.e. which are cubic or quartic in the perturbed fields) will have additional information about the physical processes at play.

A different point of view is to look at what seem to be the building blocks of large scale structures, i.e. galaxies and clusters of galaxies or, from the CDM paradigm, halos of dark matter. This more "granular" view of the distribution of matter can be used to constrain a self consistent model -- the halo model -- which quantifies the structure of the universe in terms of the density profile of individual halos, $\rho(r)$, the number density of halos of a give mass, $n(M)$ and the linear powerspectrum of fluctuations on the largest scales \citep{Cooray2002}.

While there has been tremendous progress in developing efficient quantitative methods for the non-linear regime, and they, arguably, may be of sufficient accuracy for the analysis of current data sets, much work still needs to be done to develop a fully accurate and general approach to this regime.

\subsection{$\Lambda$CDM}
Putting all the pieces together, it is possible to come up with a consistent picture of what the Universe looks like on large scales. It is accurately described by GR with a cosmological constant ($\Lambda$) making up roughly $70\%$ of the energy budget, pressureless cold dark matter (CDM) making up roughly $25\%$ of the energy budget and baryons making up $5\%$. The evidence for this model has accumulated over two decades. The first compelling evidence came from measurements of the luminosity distance through observations of distant supernovae IA \citep{Perlmutter1998,Riess1998}. Crucially, a few years later, measurements of the angular power-spectrum of the CMB from ground-based and balloon borne experiments found the first constraints on the geometry of the Universe \citep{deBernardis1999,Hanany2000,Miller2002}, ruling out curved models and singling out a flat Universe with a cosmological constant. Constraints from the abundance of clusters \citep{Bahcall2002} placed constraints on the fractional energy density in dark matter, $\Omega_M$, further cementing $\Lambda$CDM as the preferred model for describing the Universe. Subsequent cosmological observations, which we will allude to in Section \ref{LSS} have led to precision constraints (in some case at the sub-percent level) on the parameters of $\Lambda$CDM.

\section{Modified gravity and cosmology.}
\label{modgrav}
\subsection{What is General Relativity?}
Before we delve into what it means to modify gravity, let us establish what we mean by standard gravity. We take it to be Einstein's General Theory of Relativity, where the gravitational field is a metric, $g_{\alpha\beta}$ whose dynamics is described by the Einstein-Hilbert action, and which is minimally coupled to matter. In other words, gravity is encapsulated in the action
\begin{eqnarray}
\frac{1}{16\pi G}\int d^4x\sqrt{-g}(R-2\Lambda)+\int d^4x\sqrt{-g}{\cal L}_M(g_{\alpha\beta},\cdots) \label{GR}
\end{eqnarray}
where $G$ is Newton's constant, $g$ is the determinant of $g_{\alpha\beta}$, $R$ is the Ricci scalar, $\Lambda$ is the cosmological constant and ${\cal L}_M$ is the Lagrangian density for the matter and radiation fields which are minimally coupled to the metric. It is often convenient to work with the reduced Planck mass, $M^2_{\rm Pl}=1/8\pi G$. The equations of motions are then given by
\begin{eqnarray}
G_{\mu\nu}\equiv R_{\mu\nu}-\frac{1}{2}Rg_{\mu\nu}=-\Lambda g_{\mu\nu}+8\pi G T_{\mu\nu} \label{eomGR}.
\end{eqnarray}
where we have defined the Einstein tensor, $G_{\mu\nu}$.

GR is an incredibly special theory, as we hope we made clear in Section \ref{introduction}. Before we move venture into the unknown, it is useful to signal potential pitfalls and problems that arise when one tries to consider extensions.  To begin with, GR is endowed with properties that make it a {\it predictive} theory. The equations are highly non-linear and, a-priori, there is no reason to assume that it allows a well-posed initial value problem. In other words, given a set of initial conditions, is there a unique and well-defined evolution for the metric, given by equations \ref{eomGR}? It turns out that there is \citep{Choquet1969} and this means that it is possible to trust numerical solutions to the evolution equations, even in the mildly strong field regime such as, for example, in binary black hole mergers \citep{Lehner2014}. The same cannot be said for the types of extensions we will see below \citep{Allwright2018}. 

Furthermore, as we shall see, extensions to GR almost universally involve considering new degrees of freedom which, quite easily, may be unstable in one form or another. If one considers equations of motions that have higher derivatives, these may suffer from the Ostragradski instability \citep{Woodard2007}; they may also have ghosts or tachyonic instabilities that may render these theories phenomenologically unviable. Any attempt at venturing beyond GR must bear these serious problems in mind.

\subsection{Modified gravity as extra degrees of freedom.}
In the introduction, we stated Lovelock's theorem and we mentioned that it was a useful guide to how we might deform gravity {\it away} from Equation \ref{GR}. Let us now use it to explore a (by no means exhaustive) set of alternatives. For a start, let us consider adding an extra, scalar degree of freedom. One of the most iconic alternatives to GR is Jordan-Brans-Dicke (JBD) gravity \citep{JBD} in which the Einstein-Hilbert term is transformed as $\frac{M^2_{\rm Pl}}{2}R\rightarrow \varphi R$ and the scalar field is endowed with dynamics with a kinetic energy of the form $\omega(\partial_\mu\varphi\partial^\mu\varphi)/\varphi$. In practice this involves making the Planck mass (or Newton's constant) dynamical, with one free parameter, $\omega$. JBD gravity is a particular example of a scalar-tensor theory for which there is an (almost) complete description if one requires (as in Lovelock's theorem) that the equations of motion are $2^{\rm nd}$ order. Known as the Horndeski (or Covariant Galileon) action \citep{Horndeski1974,Deffayet2009,Koboyashi2011}, it is given by:
\begin{eqnarray}\label{eq:L_horndeski}
S=\int d^4x \sqrt{-g}\left\{\sum_{i=2}^5{\cal L}_i[\varphi,g_{\mu\nu}]+{\cal L}_M[g_{\mu\nu},\cdots]\right\},
\end{eqnarray}
and the building blocks of the scalar field lagrangian are
\begin{eqnarray}
{\cal L}_2&=& K ,  \nonumber \\
{\cal L}_3&=&  -G_3 \Box\varphi , \nonumber \\
{\cal L}_4&=&   G_4R+G_{4X}\left\{(\Box \varphi)^2-\nabla_\mu\nabla_\nu\varphi \nabla^\mu\nabla^\nu\varphi\right\}  , \nonumber \\
{\cal L}_5&=& G_5G_{\mu\nu}\nabla^\mu\nabla^\nu\varphi
-\frac{1}{6}G_{5X}\big\{ (\nabla\varphi)^3
-3\nabla^\mu\nabla^\nu\varphi\nabla_\mu\nabla_\nu\varphi\Box\varphi 
 \nonumber \\ & & 
+2\nabla^\nu\nabla_\mu\varphi \nabla^\alpha\nabla_\nu\varphi\nabla^\mu\nabla_\alpha \varphi
\big\}   \,.
\end{eqnarray}
We have that $K$ and $G_A$ are functions of $\varphi$ and $X\equiv-\nabla^\nu\varphi\nabla_\nu\varphi/2$, and the subscripts, $X$ and $\varphi$, denote derivatives. The four functions, $K$ and $G_A$ completely characterize this class of theories. Further extensions have been found that preserve the $2^{\rm nd}$ order nature of the equations of motion \citep{Zuma2014,Gleyzes2015a}. 

Of particular interest are theories in which additional symmetries are imposed and which will, through naturalness arguments, be protected at the quantum level. Of particular note are Galileon theories \citep{Nicolis2009} where the action is invariant under field redefinitions of the form $\varphi \rightarrow \varphi+c_\mu x^\mu +d$. Other possibilities are scale invariant theories \citep{Blas2011,GarciaBellido2011,Ferreira2016}, invariant under joint field/metric transformations of the form $\varphi\rightarrow \lambda^{-1} \varphi$ and $g_{\alpha\beta}\rightarrow \lambda^2 g_{\alpha\beta}$.

There are also other possible choices for extra fields. So, for example, one can opt for a four-vector, $A_\mu$, leading to a different class of theories. Of particular note are Einstein-Aether theories \citep{Jacobson2000} (and their generalizations --\cite{Zlosnik2007}) in which one imposes a time-like constraint, $A_\mu A^\mu+1=0$; such theories are used as models for spontaneously broken Lorentz symmetry. Other possibilities are extensions of the Maxwell-Proca action, in which the vector field is given an explicit mass \citep{Heisenberg2014}. 
It is possible to go yet further and include another tensor field, i.e. another metric $f_{\alpha\beta}$, as the extra degree of freedom. Such theories have garnered a fair amount of attention recently: by constructing a potential, $V(g^{-1}f)$ one can obtain a theory of a massive graviton which isn't plagued by instabilities, most notably ghosts \citep{Hinterbichler2012,deRahm2011,Hassan2012}. There are, of course, a myriad of possibilities, adding multiplets of fields or a combination of fields with different spins \citep{Clifton2012}.

\subsection{Alternatives: higher derivatives, higher dimensions and non-locality.}
Instead of explicitly adding extra fields, it is possible to do so implicitly in a variety of different ways. One of the simplest ways is to replace $R$ in the Einstein-Hilbert action by a function of other curvature invariants, $R\rightarrow f(R,R^2, R_{\alpha\beta}R^{\alpha\beta}, R_{\alpha\beta\mu\nu}R^{\alpha\beta\mu\nu},\cdots)$ where $R_{\alpha\beta}$ and $R_{\alpha\beta\mu\nu}$ are the Ricci and Riemann tensors. In doing so, one introduces higher derivative terms in the action which is tantamount to adding new solutions to the equation of motion or, equivalently, adding new degrees of freedom. This can be made clear in the particular case where the $f$ depends solely on the Ricci scalar \citep{Sotiriou2010}; by defining a new scalar field, $\varphi\sim \ln df/dR$, the theory can be re-expressed as scalar-tensor theory, a subset of the theories discussed above.

One approach which emerges from attempts at unification, such as string theory, is to enlarge the number of dimensions. A notable example is the Kaluza-Klein model \citep{Clifton2012} in which there is one extra dimension and the dynamics is given by the 5-dimensional Einstein-Hilbert action for the metric $g_{AB}$ where $A$, $B$ take values from $0$ to $4$. By redefining the metric as $g_{AB}=e^{\varphi/\sqrt{3}}g_{\mu\nu}+e^{-\varphi/\sqrt{3}}A_\mu A_\nu$ when $A$, $B$ take values from $0$ to $3$, $g_{\mu4}=e^{-\varphi/\sqrt{3}}A_\mu$ and $g_{44}=e^{-\varphi/\sqrt{3}}$,  it is possible to trivially equate a higher dimensional theory with a lower dimensional theory with extra fields (albeit in which the fields have an extra coordinate dependence on $x^4$). By compactifying the extra dimension or by confining the dynamics to a 3-space \citep{Arkani1998,Randall1999}, it is possible to obtain an approximate, but not exactly, $3+1$ universe with modified gravitational dynamics (given by extra fields). A particularly popular model -- the DGP model -- was proposed in \citep{Dvali2000} of a three dimensional "brane" living in a five dimensional space. Through a judicious choice of the brane geometry and tension, vis-a-vis the background space-time it is possible to obtain a "self"-accelerating Universe which mimics one with a cosmological constant (or with dark energy). Furthermore, deviations from GR emerge on large scales (and not on small scales as one expects in "conventional" higher dimensional universes).

The final, more exotic, direction in which we can take this is by considering non-locality \citep{Deser2007,Maggiore2014}. It is possible to, for example, include terms in the action of the form $R\times R/\Box$ where the inverse D'Alembertian can now be interpreted as a Greens function (which is by definition non-local). Interesting effects arise in a cosmological context due to the non-locality in time and it can be shown that such theories are equivalent to  scalar tensor theories by replacing $\Box\varphi\equiv R$ and imposing strict constraints on the initial conditions in such a way as to kill all spurious modes arising from the homogenous solutions to this equation. 

\subsection{The speed of gravitational waves and GW170817}

Before I delve into the cosmological regime of gravitational theories, it is important to briefly highlight a recent event. The discovery of a binary neutron star (BNS) merger with both gravitational waves and a multitude of electromagnetic signals has heralded the advent of {\it multi-messenger} astronomy \citep{Abbott2017}. One of the remarkable bi-products has been the ability to place stringent constraints on the speed of gravitational waves: it is within $10^{-15}$ of the speed of electromagnetic waves. This fact allows us to severely restrict the range of allowed modified gravity theories that have an impact on cosmological scales \citep{Baker2017,Creminelli2017,Ezquiaga2017} -- the extra degrees of freedom will serve as a medium through which the gravitational waves propagate and, depending on their coupling to gravity, may affect their speed. It has been shown that, for example, scalar-tensor theories are strongly constrained by the multi-messenger detection of the BNS, as is a broad swathe of vector-tensor theories.

\subsection{General theories of gravity on cosmological scales.}
I briefly covered a variety of ways in which one can extend GR and we have seen that, in general, it can be seen as adding extra degrees of freedom. But we have also seen that the landscape of possible extensions is vast and resides in large functional space of possibilities. This will give pause for thought: given that there is a restricted amount of observational data, the task of comprehensively constraining GR seems difficult at best. The key to progress is to realize that, in this endeavour, one should focus on specific regimes in which certain characteristics of the theory may simplify. Indeed, this is standard practice in high-energy physics, when one focuses on the relevant aspects of a theory at a given energy scale (for example, at the beam luminosity of a particular accelerator). Furthermore, this approach is already standard practice in, for example, the analysis of gravitational data on the scale of the Solar System. There, an efficient formalism for tackling weak field, non-relativistic systems -- the Parametrized Post Newtonian (PPN) formalism -- has been used to characterize (non)-deviations from GR of various forms \citep{Thorne1971}. It would make sense, then, to apply the same rationale to cosmology.

At its most basic level, cosmology is about the expansion of the Universe, i.e. a detailed description of the time evolution of the scale factor, $a(t)$. As we have seen in GR, this allows us to disentangle the different constituents of the Universe, by constraining the $\Omega_X$ in the Friedman equation. Of particular interest is the fact that we may be able to constrain the equation of state of the dark energy component, $w_{\rm DE}$. In other words, observables such as $H(z)$ or $D_A(z)$ can be converted into constraints on $w_{\rm DE}(z)$ which in turn can be used to extract the fundamental nature of the dark energy (i.e. if it is a cosmological constant, if it is a scalar field with a particular form of the potential, $V(\varphi)$, etc).

How does the situation change if one now wishes to constrain gravity itself? Unfortunately, not very much. To see why this is so, we need to realize that {\it any} modification to GR can be written in the form
\begin{eqnarray}
G_{\mu\nu}=-\Lambda g_{\mu\nu}+8\pi G T_{\mu\nu}+U_{\alpha\beta}
\end{eqnarray}
where $U_{\alpha\beta}$ can be a functional of $g_{\alpha\beta}$, the matter fields and any extra degree of freedom (such as scalar, vector and tensor fields). If we now impose homogeneity and isotropy, we have that $U_{\alpha\beta}={\rm diag}(8\pi G \rho_U,8\pi G P_U\delta_{ij})$ where  $(\rho_U,P_U)$ are two free functions of time; note that we have chosen a notation that shows that, at the background level,  modifying GR is essentially equivalent to adding a new energy-momentum tensor, which can also be characterized by an equation of state, $w_{\rm U}$. This means that, at the background level, modifying GR is indistinguishable from adding in a dark energy fluid \citep{Kunz2012} and it is impossible to distinguish between the two with observables such as $H(z)$ or $D_A(z)$.

Given the paucity of information in the background, we need to delve a level deeper in complexity and look at linear perturbations of homogeneous and isotropic background space-time. As we saw, in GR, this leads to a series of  $2^{\rm nd}$ order coupled differential equation (as well as constraints) for the perturbed variables $\delta_X$, $\theta_X$, $\Phi$ and $\Psi$. If we now go beyond GR, but remain in the arena of linear perturbations, we can easily guess what the structure of these "extended" equations might be \citep{Baker2012}. They will be linear and (for now) $2^{\rm nd}$ order but they will include new, extra degrees of freedom. So, for example, if the extra degree of freedom arises from a scalar field, $\varphi$, we will now need to include its perturbation, $\delta\varphi$.

Let us now take two of the perturbed Einstein field equations and generalize them. We can modify Equation \ref{NP} to
\begin{eqnarray}
-k^2\Phi=4\pi G \Delta\rho+A_0k^2\Phi+F_0k^2+\alpha_0k^2\delta\varphi+\alpha_1k\delta{\dot \varphi}+\cdots\Gamma \label{NPPPF}
\end{eqnarray}
and Equation \ref{Slip} to
\begin{eqnarray}
k^2(\Phi-\Psi)=D_0k^2\Phi+D_1k{\dot \Phi}+K_0k^2\Gamma+K_1k{\dot \Gamma}+\epsilon_0k^2\delta\varphi+\epsilon_1k \delta{\dot \varphi}+\epsilon_2 \delta{\ddot \varphi}\label{SlipPPF}
\end{eqnarray}
where $k\Gamma=(\dot{\Phi}+{\cal H}\Psi)$ and $A_0$, $D_0$, $D_1$, $F_0$, $K_0$, $K_1$, $\alpha_0$, $\alpha_1$, $\epsilon_0$, $\epsilon_1$ and $\epsilon_2$ are free functions of time and $k$; setting these free functions to zero recovers GR but, in principle, one can span the most general class of gravitational theories which involve an extra scalar field. The generality of this approach is appealing yet one is clearly faced with a challenge: is it feasible to, with a finite amount of cosmological data, constrain these multiple functions of time and space?

We have been working at the level of the equations of motion but, in almost full generality, alternative theories to GR will obey an action principle. This means that it would make more sense to extend the action itself. We have already seen that at the full, non-linear level, with the Horndeski action presented in Equation \ref{eq:L_horndeski}. But if one limits oneself to the most general theory at the {\it linear} level, the exercise simplifies greatly: one simply has to construct the most general action, quadratic in the perturbation variables and which satisfies any desired symmetry principle, the most important of which is linearized diffeomorphism invariance, i.e. invariance under transformations of the type $x^\alpha\rightarrow x^\alpha+\xi^\alpha$. 

There are a few ways in which one can go about constructing the extend linear action. A systematic way is to write the most general quadratic action and then impose symmetries to reduce the number of terms \citep{Battye2011,Lagos1,Lagos2,Lagos3}. This has been shown to be general and applicable to any type of extra degree of freedom: scalars, vectors and tensors. If one restricts oneself to the case where the extra degree of freedom is a scalar field, a simpler, more efficient approach can be taken with what has been dubbed the Effective Field Theory (EFT) approach \citep{Creminelli2009,Gubitosi2012,Bloomfield2012,Gleyzes2013,Gleyzes2015b}: starting with a general action which only depends on metric perturbations, one then performs a linearized diffeomorphism transformation, in which $\xi^0=\delta\varphi$ and, in such a way one generates the most general linearized scalar-tensor action.

It turns out that, in the case of scalar-tensor theories, there is a straightforward approach of obtaining the most general action for linear perturbations: one takes the Horndeski action of Equation \ref{eq:L_horndeski} and Taylor expands to $2^{\rm nd}$ order in the perturbations. If one does that, one finds that, instead of depending on a multitude of free functions of time and scale, as we saw above, the resulting action merely depends on four functions of time, $\alpha_A(t)$ with $A=K,B,M,T$ which can be directly related to the original action \citep{Bellini2014}. These functions are
\begin{eqnarray}
M^2_*&\equiv&2\left(G_4-2XG_{4X}+XG_{5\varphi}-{\dot \varphi}HXG_{5X}\right) , \nonumber \\
HM^2_*\alpha_M&\equiv&\frac{d}{dt}M^2_* , \nonumber \\
H^2M^2_*\alpha_K&\equiv&2X\left(K_X+2XK_{XX}-2G_{3\varphi}-2XG_{3\varphi X}\right) \nonumber \\ & &
+12\dot{\varphi}XH\left(G_{3X}+XG_{3XX}-3G_{4\varphi X}-2XG_{4\varphi XX}\right) \nonumber \\ & &
+12XH^2\left(G_{4X}+8XG_{4XX}+4X^2G_{4XXX}\right)\nonumber \\ & &
-12XH^2\left(G_{5X}+5XG_{5\varphi X}+2X^2G_{5\varphi XX}\right)\nonumber \\ & &
+14\dot{\varphi}H^3\left(3G_{5X}+7XG_{5XX}+2X^2G_{5XXX}\right)\nonumber , \\
HM^2_*\alpha_B&\equiv&2\dot{\varphi}\left(XG_{3X}-G_{4\varphi}-2XG_{4\varphi X}\right) \nonumber \\ & &
+8XH\left(G_{4X}+2XG_{4XX}-G_{5\varphi}-XG_{5\varphi X}\right)  \nonumber \\ & &
+2\dot{\varphi}XH^2\left(3G_{5X}+2XG_{5XX}\right) \nonumber , \\ 
M^2_*\alpha_T&\equiv&2X\left[2G_{4X}-2G_{5\varphi}-\left(\ddot{\varphi}-\dot{\varphi}H\right)G_{5X}\right] \,. \label{alphas}
\end{eqnarray}
Each of these functions are linked to specific physical properties of the theory: $M^2_*$ and $\alpha_M$ are related to time variations in the background Newton's constant, $\alpha_K$ to the generalized canonical kinetic term of simple DE models, $\alpha_B$ quantifies kinetic mixing between $\varphi$ and the scalar perturbations of the metric and $\alpha_T$ is associated to modifications to the speed of propagation of tensor modes. 

The fact that general theories of gravity can, at the level of linearized cosmological perturbations, be completely characterized by a handful of free functions is true in general, including, for example, the case where the extra degrees of freedom are 4-vectors or tensors. And it gives hope to the idea that it might be feasible to constrain gravity on large scales. 
\subsection{The quasi-static regime.}
It turns out that there is a further simplification if one restricts oneself to sufficiently sub-horizon scales, i.e. scales in which $k/(aH)\sim k\eta\gg 1$ where $ad\eta=dt$ allows us to define the conformal time $\eta$. Known as the "quasi-static" regime (the cosmological equivalent of the Newtonian limit), this covers scales up to hundreds of Megaparsecs and thus encompasses most of the current and future large scale structure surveys under consideration. In that regime, it is an excellent approximation to effectively freeze out the dynamics of the extra degree of freedom; for example, in the case of a scalar field, this means approximating $\Box \varphi\propto {\rm "source"}$ by $-\nabla^2 \varphi\propto {\rm "source"}$. Replacing the extra degree of freedom in the generalized Einstein field equations leads to modified equations \ref{NP} and \ref{Slip}:
\begin{eqnarray}
-k^2\Phi&=&4\pi G \mu \Delta\rho \label{NPMG} \\
\Phi&=&\gamma \Psi \label{SlipMG}
\end{eqnarray}
where $\mu$ is the modified, effective Newton's constant (and can be rewritten as $\mu=G_{\rm eff}/G_0$) and $\gamma$ is often called the gravitational slip \citep{Hu2007,Bertschinger2008,Amin2009}. The quasi-static parameters $\mu$ and $\gamma$ are generally functions of time and scale, $k$; typically the scale dependence either kicks in on very small scales (i.e. deeply in the non-linear regime) or naturally, close to the cosmological horizon when the quasi-static approximation breaks down. This means, that, effectively, when constraining theories of gravity on cosmological scales, one is constraining two free functions of time. An approximate approach is to extend this parametrization to all scales, including super-horizon scales, greatly simplifying any attempts at cosmological parameter estimation.

Note that, it is customary to assume that the quasi-static regime is synonymous with the use of $\mu$ and $\gamma$ on subhorizon scales -- we will do so in this review. But we need to point out that this is not entirely general and that in Beyond Horndeski theories \citep{Mancarella2017}, there will be additional terms in equations \ref{NPMG} and \ref{SlipMG} proportional to the matter velocity which also have to be taken into account; this means one would need to extend the number of functions in the quasi-static regime from two to four. 

Over the past few years, quite a few variants of $\mu$ and $\gamma$ have been proposed, it is worth our while to briefly cover them. Some authors choose the following notation 
\begin{eqnarray}
-k^2\Psi&=&4\pi G \mu \Delta\rho \label{NPMGn} \\
\Phi&=&\eta \Psi \label{SlipMGn}
\end{eqnarray}
If we redefine the $\mu\rightarrow {\tilde \mu}$ we have ${\tilde \mu}=\mu/\gamma$ (where $\mu$ is defined in Equation \ref{NPMG}) and $\eta=\gamma$. ${\tilde \mu}$ is of particular use as (we shall see) it sources the growth rate of structure. In the same way, we can define a parameter that sources weak lensing
\begin{eqnarray}
\Sigma=\mu\left(1+\frac{1}{\gamma}\right)={\tilde \mu}(1+\eta)
\end{eqnarray}
Some authors often define $G_M$ and $G_L$ (where $M$ stands for "Matter" and $L$ stands for "Light") such that
\begin{eqnarray}
G_M={\tilde \mu} \ \ \ G_L=\Sigma.
\end{eqnarray}

The quasi-static parameters can be directly connected to more fundamental parameters in underlying theories. For example, in the case of scalar-tensor theories arising from the Horndeski action, these parameters have a specific structure \citep{DeFelice2011}
\begin{eqnarray}
\mu&=&\frac{1+h_1k^2}{h_2+h_3k^2} \nonumber \\
\gamma&=&\frac{h_2+h_3k^2}{h_4+h_5k^2} \nonumber
\end{eqnarray}
where $h_i$, $i=1,\cdots 5$ are functions of time and assembled from the 4 $\alpha_X$ parameters from equations \ref{alphas}; thus constraints on the quasi-static parameters lead to constraints on slices of the functional space of free parameters, $\alpha_X$. And vice versa, by restricting one self to scalar tensor theories, one can establish a restricted set of priors on $\mu$, $\gamma$ or $\Sigma$ \citep{Pogosian2016,Peirone2018}

\subsection{The phenomenology of modified gravity.}
The dynamics of linear perturbations in modified versions of GR is sufficiently well understood that there are now codes that accurately model the evolution of perturbations in such theories, calculating cosmological observables. The basic framework is a code that integrates the perturbed Einstein field equations, the coupled set of matter (and other fluid like) conservation equations and the Boltzman equations for radiation and neutrinos. The main, general purpose, Einstein-Boltzman solvers are EFTCAMB \citep{Hu2013} and MGCAMB \citep{Hojiati2011} and ISITGR \citep{Dossett2011}  (derived from CAMB \citep{Lewis1999}) and HiCLASS \citep{Zuma2017} derived from CLASS \citep{Blas2011} although over a dozen more specific Einstein-Boltzman solvers have been written. Most of these codes have been cross calibrated and agree at the sub-percent level \citep{Bellini2017}.

While we have focused on how we can accurately model general deviations from GR, we now discuss how to constrain these deviations. The most fundamental constraint will come from measurements of the growth rate: from redshift space distortion (RSD), peculiar velocities, the integrated Sachs-Wolfe effect and from tomographic measurements of weak lensing. All of these will constrain the time evolution of $\delta_{\rm M}$ through $f$ or $f\sigma_8$. To see how the growth rate is affected by deviations from GR, it is instructive to focus again on the quasi-static regime \citep{Baker2014}. There, the evolution equation is now given by
\begin{eqnarray}
\frac{df}{d\ln a}+q(a)f+f^2=\frac{3}{2}\Omega_{\rm M}(a)\frac{\mu}{\gamma}
\end{eqnarray}
where $q(a)=\frac{1}{2}[1-3w_{\rm U}(a)(1-\Omega_{\rm M}(a))]$. We can clearly see that the growth rate will be affected by both modifications to the background (which is encapsulated here in the equation of state of the effective fluid driving the acceleration, $w_{\rm U}$) and the modified perturbed equations through the quasi-static parameters. As a result there will be a degeneracy between $w_{\rm U}$ and $\mu/\gamma$ \citep{Simpson2010} which can only be broken with measurements of cosmological distances. A particular parametrization of the growth is often used, of the form
\begin{eqnarray}
f=\Omega_M^{\gamma_G}
\end{eqnarray}
where $\gamma_G$ is approximately a constant for any specific theory over a narrow range of redshifts \citep{Linder2005}. For $\Lambda$CDM, it can be shown that $\gamma_G=6/11$.

Weak lensing will also probe the growth rate -- via the modified Newton-Poisson equation it measures the change in the density contrast \citep{Amendola2008}. But it goes further; if we look at equation \ref{lensing} we can see that it depends on both $\Phi$ and $\Psi$. If we now focus on convergence we have that $\kappa\propto \int d\chi \nabla^2(\Phi+\Psi)\sim \int d\chi \nabla^2(1+\frac{1}{\gamma})\Phi\sim 
\int d\chi \frac{3}{2}(aH)^2\Omega_{\rm M}\mu(1+\frac{1}{\gamma})\delta_{\rm M}$ where we used the Friedman equation to simplify the result. In other words, as well as depending on the growth rate (via $\delta_M$) it probes one of the derived quasi-static parameters described above: $\Sigma$. 

Extending the set of cosmological parameters to include extra parameters (such as $\mu$ and $\gamma$) may lead to degeneracies -- we have already seen such a case between $w_{\rm U}$ and $\mu/\gamma$ in the case of the growth rate. A particularly problematic parameter is galaxy bias, i.e. the mapping between the galaxy density contrast, $\delta_{\rm G}$ and the underlying matter density contrast, $\delta_{\rm M}$. A simple approximation is that $\delta_{\rm G}\simeq b_{\rm G}\delta_{\rm M}$ where $b_{\rm G}$ is scale and time dependent (although there are now, far more sophisticated models of bias \citep{Desjacques2016}). Given that many direct probes of the growth rate are in terms of the galaxy distribution, one often finds that resulting constraints are on $f/b_{\rm G}$. 

A proposal to circumvent the dependence on bias is to consider combinations of statistics that, to some extent, "divide out" the dependence on bias \citep{Zhang2007}. A notable example is
\begin{eqnarray}
E_G(R)\propto\frac{\xi_{\rm GL}(R)b_{\rm G}}{\xi_{GG}(R)}
\end{eqnarray}
where $\xi_{\rm GG}(R)$ and $\xi_{GL}(R)$ are the 2-D galaxy-galaxy and galaxy-lensing correlation functions and $b_G$ is an independent measurement of the galaxy bias of the same sample. This statistic was originally formulated in Fourier space but is implemented in real space where a number of practical issues need to be considered \citep{Leonard2015}. Alternatively one can consider the full suite of correlation functions (including the lensing-lensing correlation function) which, combined, can break the degeneracy between $b_{\rm G}$ and other parameters.

\section{Fifth forces and gravitational screening.}
\label{screening}
\subsection{Gravitational regimes and the observational desert.}
In the previous section we focused on what one might consider to be a "clean" regime, where dynamics is well described by linear cosmological perturbation theory, and showed how it was relatively straightforward to parameterise general theories of gravity. This regime benefits from the fact that the dynamics is perfectly understood and it is possible to solve the evolution equations with arbitrarily high accuracy. As we saw in Section \ref{cosmology}, it is where modern cosmology has had a resounding success. But it is also a regime with severe limitations: the larger the scales we have, the less statistical power to constrain parameters. We observe one universe and hence there is a finite amount of information to be extracted from its observation; this cosmic variance is largest for the largest scale modes. It would be desirable to look at smaller scale modes but, as we have seen, non-linear physics (both gravitational and baryonic) play an unavoidable role there.

\begin{figure*}[h]
\includegraphics[width=3in]{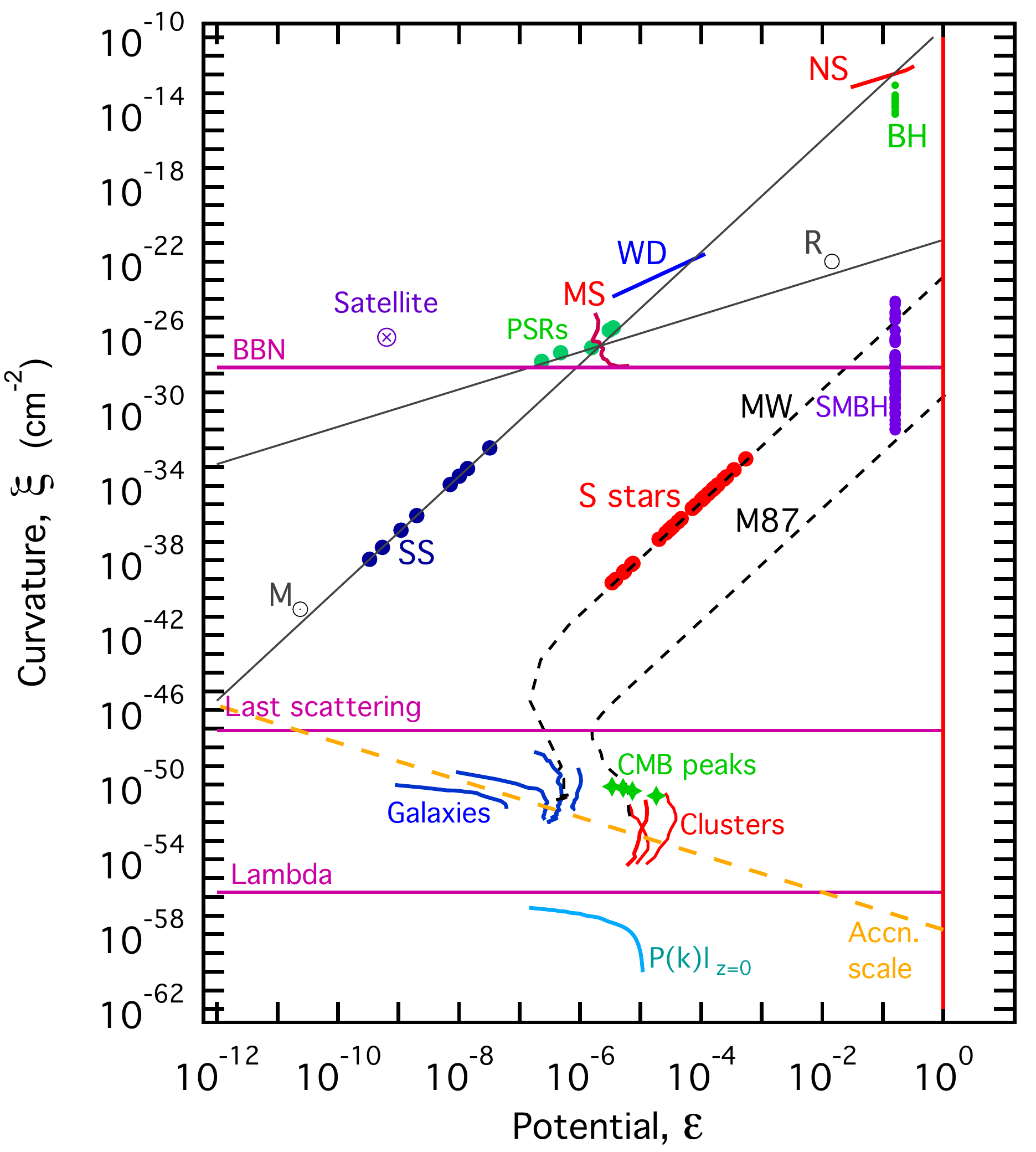}
\caption{Gravitational systems as a function of the typical values of their gravitational potential and curvature \citep{Baker2015}. The Solar System is on the middle left, while black holes are on the top right. Cosmological structures are at the bottom.}
\label{BPS}
\end{figure*}

There is another way of looking at the large scale, cosmological regime in comparison to other regimes \citep{Baker2015a} and this is best illustrated in Figure \ref{BPS}. On the horizontal axis one has the typical gravitational potential, $\Phi$ (in this plot labelled by $\epsilon$) of a system while on the vertical axis one has the typical curvature  (labelled by $\xi$), here expressed in terms of the Kretschmann scalar (i.e. a quadratic invariant constructed from the Riemann tensor). Scattered on the plot are a number of different systems (or regimes in which gravity can be constrained), from the Solar System and pulsars to black hole mergers and imaging. Cosmology stands apart in a regime of low curvature.

A striking feature of Figure \ref{BPS} is that there seems to be a "desert", a dearth of observation between cosmology and other probes, for values of $\xi$ between $10^{-50}$ and $10^{-42}$. The typical systems found in this desert are galaxies and clusters where the interplay of non-linear gravitational and baryonic physics as well as highly non-linear feedback mechanisms (such as supernovae and Active Galactic Nuclei - AGNs) make them particularly hard to model. Indeed we can see this desert as the transition between two easy regimes: the high curvature part where one is studying isolated systems and the low curvature part where linear theory on a homogeneous background applies.

There is another interesting feature that can be observed. In the high curvature regime, there are strong constraints on gravity - in the case of the Solar System and millisecond pulsars, exceptionally so. One is guaranteed to have modifications to GR in the top right-hand side as one delves into the regime where gravity becomes much stronger. But the current quest for modifications are in the bottom half of the regime where, we argued in the introduction, it is somewhat unnatural for deviations to GR to arise. If it turns out that GR is strictly true in the top half of the plane and violated in the bottom half of the plane, it would seem that some mechanism is shielding regions of high potential from deviations, i.e. from extra fifth gravitational forces and that the transition occurs somewhere in the desert.

\subsection{Gravitational screening.}
While the description above is somewhat qualitative, it turns out that there are mechanisms of what is known as {\it gravitational screening} which can restore GR in regions of high density, potential or curvature but which allow for the emergence of fifth forces elsewhere \citep{Joyce2014}. A useful, alternative, way of thinking about screening is to realize that a fifth force is always associated to a "charge" and that the strength of the fifth force is proportional to the charge. One can then envisage a mechanism such that  in the screened regime the charge is surpressed and the fifth force is absent while in the unscreened regime, the force has full effect. 

To understand how screening can work, let us revisit one of the simplest scenario, a non-minimally coupled scalar field with a potential \citep{Khoury2004,Burrage2017}. It is always possible to transform such a system into a frame in which the scalar field becomes minimally coupled to gravity but non-minimally coupled to the matter sector. In the simplest case, the background scalar field obeys an equation of the form
\begin{eqnarray}
\Box \varphi+\frac{dV_{\rm eff}}{d\varphi}=0
\end{eqnarray}
where the effective potential, $V_{\rm eff}=V-{\bar \rho}\varphi$, i.e. the minimum of the potential will depend on the local matter density. If we solve for $dV_{\rm eff}/{d\varphi}(\varphi_0)=0$, we can then expand $\varphi=\varphi_0+\delta\varphi$ and we find that, in the quasi-static regime, the perturbations in the scalar field obey an equation of the form
\begin{eqnarray}
\nabla^2 \delta\varphi+m^2(\varphi_0)\delta\varphi\propto "{\rm source}" \nonumber
\end{eqnarray}
where $m^2(\varphi_0)=d^2V_{\rm eff}/{d\varphi}^2$. If we assume a localized mass, $M$ this means that
\begin{eqnarray}
\delta\varphi= \beta \frac{Me^{-m(\varphi_0)r}}{r} \label{screen}
\end{eqnarray}
where $\beta$ is a constant of proportionality set by the coupling between the scalar field to matter and we see that $m(\varphi)$ functions as a cutoff, setting the range of the fifth force. Furthermore, the cut-off depends on the environment so that, if we choose a $V\propto \varphi^{-n}$, we have that for large $\rho$, $m(\varphi_0)$ is large and the force is short range but if $\rho$ is small, $m(\varphi_0)$ will be small and the force can be long range. Thus we have that the fifth force can strongly satisfy local constraints yet have an impact on cosmological scales.

We have discussed one particular type of screening, known as {\it Chameleon} screening. Interestingly, it has been shown that, if Chameleon screening is to arise in a cosmological context, it has to be on scales less than a Megaparsec, i.e. deep in the non-linear regime; furthermore, the screening scalar field cannot be responsible for accelerated expansion directly through gravitational effects and only as a conventional source of dark energy \citep{Wang2012}. There are, however, other forms of screening that we should mention. The first one arises in theories in which the scalar mediator of the fifth force has a non-trivial kinetic term, as in, for example bimetric theories of massive gravity or scalar-tensor theories constructed from shift-symmetric term -- Galilleons \citep{Nicolis2009} -- or bimetric theories such as massive gravity. Known as the {\it Vainshtein} mechanism \citep{Vainshtein1972}, the modified kinetic term now transforms the overall strength of the fifth force, the $\beta$ in equation \ref{screen}, into a function of $\varphi_0$ which in turn can depend on the source and environment. Typically one find that near very massive objects, $\beta\rightarrow 0$ while around lighter or more diffuse objects, $\beta$ can be non-negligible. Another variant is {\it Symmetron} \citep{Hinterbichler2010} where an environmental dependence of the coupling between the scalar field and matter leads to a similar effect on $\beta$.

\subsection{The phenomenology of screening.}
Screening adds more complexity on to what is already a particularly difficult regime to model. While the payoff for looking at smaller scales is greatly enhanced statistical power (there are more independent samples on smaller scales), one is faced with the non-linear aspects of gravity which are far less well understood than the clean linear regime discussed in the previous section. Furthermore, baryonic physics can now have a significant impact on attempts at making predictions.

There have been some attempts at developing analytic techniques for studying the impact of screening on the non-linear regime. Working in the quasi-static regime, it is possible to extend or modify higher order perturbation theory methods to included the effect of fifth forces \citep{Koyama2009,Cusin2018} although the effectiveness of such methods has been somewhat limited. More progress has been made on developing enhanced N-body codes which incorporate the effect of fifth forces \citep{Winther2015}. There, again using the quasi-static approximation, it is possible to add corrections to the gravitational forces which captures the effect of environment dependent fifth forces. Furthermore, it is possible to couple the modified gravitational evolution with hydrodynamic equations to allow a first foray into the impact of baryonic physics and how it feeds back into gravitational collapse.

Currently, the main codes used for studying the non-linear regime and its interplay with baryons are {\tt ECOSMOG} \citep{Li2012} and {\tt ISIS} \citep{Llinares2014} which are modified versions of {\tt RAMSES} \citep{Teyssier2002},  an adaptive-mesh-refinement code, and  {\tt MG-GADGET}  \citep{Puchwein2013} which is a modified version of {\tt GADGET} \citep{Springel2005}, an SPH code. There have also been attempts at writing faster N-body solvers with fifth forces, based on particle-mesh (PM) solvers, most notably modified versions of {\tt COLA} \citep{Valogiannis2017,Winther2017}. As with the Einstein-Boltzman solvers, there has been a concerted effort to cross-calibrate various codes so that they agree for a range of models on a range of scales \citep{Winther2015}. 

\subsection{Observational consequences.}
If gravitational screening is responsible for the qualitative difference between different regimes, then one needs to take a more nuanced view when undertaking a statistical analysis of large scale structure. In particular, one needs to look at the bigger picture, not only focusing on strictly local values of correlation functions or number counts, but including information about the environment in which these quantities are being estimated. One approach is almost geographic in nature: to construct a {\it gravitational map} of the Universe which allows us to identify regimes in which systems may be unscreened.

There have been attempts at constructing a gravitational map of the local Universe (i.e. which extends out to approximately $200h^{-1}$ Mpc) \citep{Cabre2012, Desmond2018a}. The basic idea in \citet{Desmond2018a} is to use a catalogue of galaxies which is as complete as possible in that region -- the 2M++ catalogue. One then uses abundance matching to identify the possible dark matter halos associated with those galaxies and to fill in for any smaller halos which will not have been picked up by the survey. Finally, one uses a constrained realization of the long wavelength modes to fill in the rest of the dark matter on large scales. From the combination of all these terms it is possible to construct a map of the gravitational potential, tidal forces and local curvature. It is also possible to model the uncertainty in all the steps leading to an overall local error budget on the map.
This map can then be used to distinguish between the regions that are unscreened and screened.

Once one can identify qualitatively different regions, it is then possible to look for signatures of the unscreened phase, i.e. for evidence of the unsupressed fifth force associated with deviations from GR. The way to go about doing this is to try and identify objects which might behave differently in unscreened environments: some object will either be sufficiently dense, compact or large that they will self screen themselves (i.e. their fifth force charge will be effectively zero) while others will be too diffuse or light to be self screened. The hall mark of the fifth force in unscreened regions will be what seem like effective violations of the equivalence principle.

In this context, there are a number of interesting effects that one might look for \citep{Jain2011}. For a start, the different components of a galaxy will behave in different ways. For example, in a theory with chameleon (or symmetron screening) the stars in a galaxy will self-screen while the gas will be too diffuse and will not be screened. The net result is that there should be an offset between the optical and radio components of a galaxy if there is screening. By looking for this effect it should be possible to place constraints on the amplitude and range of the a screened fifth force.

Another possibility is by looking at how a galaxy is deformed as it sits in a dark matter halo. The stellar disk might lag behind the halo centre inducing a gradient in the gravitational potential. As a result the galaxy will be warped into a cup shape where the amount of deformation will depend on the strength of the fifth force. The effect will be strongest if the disk is completely perpendicular to the external unscreened force field. 

A more straightforward effect will be that the unscreened component of the galaxy -- the gas -- will have a faster rotation curve relative to the stellar disk as a result of the additional fifth force having the same direction as the normal gravitational force. Furthermore, in the case of edge-on infall, the stellar and gas disks (and rotation curves) will have greater differences or assymetries with the edges which are closest to the centre of the halo being more compressed than the far side.

Finally, a similar and striking signature may be the relative offset of the galaxy centroid and the super massive black hole residing at its centre \citep{Hui2012}. If a galaxy falls in an unscreened potential, much of the galaxy will feel a fifth force while the black hole, as a result of the no-hair theorem, will be shielded and won't be subjected to it. 

Another line of attack is to roll out suitably tailored versions of the usual n-point statistics which are used in standard cosmological analysis. In particular, it makes sense to correct estimators so that they downweight screened regions and enhance unscreened regions \citep{White2016,Valogiannis2018}. In the case of chameleon screening, this can easily be achieved by downweighting regions of high density, by "clipping" high density peaks or by introducing some form of nonlinear function of the density field ${\tilde \delta}[\delta_M]$ such that high density values are damped. Examples of such transformations are ${\tilde \delta}=\ln(\delta+1)$ and ${\tilde \delta}=[(\rho_*+1)/(\rho_*+\rho_m)]^p$ where $\rho_*$ and $p$ are free parameters.

An approach that has been gaining traction is to focus on what are, for sure, the most unscreened parts of the Universe: voids. Voids have a scale of about $20-100h^{-1}$ Mpc and are typically surrounded by outflows; they are seen as a new and interesting arena in which to do cosmology, parts of the density field where many of the linear and very mildly linear results still hold. Furthermore, with the vast increase in the size and depth of cosmological surveys, it is now possible to construct meaningful statistics with void catalogues in the same way as one does with galaxy clusters. And given that voids are particularly prone to be affected by unscreened fifth forces, it makes sense to focus on them if one wants to test gravity.

As yet, the exact definition of a void has not yet been completely established and agreed on \citep{Sutter2012,Nadathur2015}. A number of different factors come into play, from dimensionality to the type and number density of the tracers used, as well as, operationally, the type of void finding algorithm that is used. As a result, void catalogues can be quite different from each other and the resulting statistics can disagree. 

Nevertheless, if one settles on one particular catalogue (or better, considers a range of catalogues) one can proceed to assess a number of statistics. In particular, one looks at the usual statistics applied in cosmological analysis except now in this very particular environment. So, for example, one can assess lensing of background galaxies by the void \citep{Barreira2015,Melchior2014,Baker2018}. One can also measure redshift space distortions in the void and compare them to measurements outside the void \citep{Hamaus2015}. Another approach is to cross correlate the void profile with the CMB \citep{Cai2014} in the same direction and in doing so, pull out the integrated Sachs-Wolfe term (the last part of equation \ref{CMBeq}). Finally, one can approach voids in the same way one approaches clusters and quantify their abundance as a function of scale and redshift \citep{Clampitt2012}.

\section{Constraints from Large Scale Structure.}
\label{LSS}
\subsection{The evidence for $\Lambda$CDM}
We seem to be living in charmed times, at least from the point of view of cosmological observations. With observations of the CMB -- with the COBE, WMAP and PLANCK satellite missions and a host of ground-based experiments -- complemented by the burgeoning precision probes of LSS, we now have an accurate mathematical model of the origin and evolution of the Universe in which the six parameters that govern it have been measured to exquisite precision \citep{Aghanim2018}. Based on GR with a cosmological constant (and described in Section \ref{cosmology}), $\Lambda$CDM fits almost all the data with remarkable consistency. 

The observational apparatus that we have to support $\Lambda$CDM is formidable. As well a the CMB experiments, the main observations to contribute to cosmological constraints are: spectroscopic surveys -- BOSS \citep{Alam2016} and WiggleZ \citep{Blake2011} -- measuring galaxy spectra (i.e. redshifts) and angular positions; photometric and weak lensing surveys -- KIDS \citep{Hildebrandt2016}, HSC \citep{Mandelbaum2018}  and DES \citep{Abbott2018}  -- measuring galaxy images and shapes, angular positions and a crude measurement of redshifts. There are preliminary attempts at line-intensity mapping (for example measuring the 21 cm spin flip transition of neutral hydrogen) but nowhere near at the level required to do cosmology \citep{Switzer2013}.

Just to belabour the point, it is useful to see how well some of the fundamental parameters of $\Lambda$CDM are currently constrained \citep{Aghanim2018}. The initial conditions of fluctuations are characterized by a spectal scalar index, found to be $n_S=0.965\pm0.004$, i.e. almost but not exactly scale invariant; the difference from scale invariance is $8-\sigma$, consistent with (and, indeed, some might say strongly indicative of) a period of primordial inflation. This constraint is supplemented with an upper bound on the tensor to scalar ratio (i.e a measure of the amount of gravitational waves left over from inflation) of $r<0.07$. Another notable constrain is on the geometry of the universe: the current, tightest constraint is $\Omega_K=-0.001\pm0.002$. The Hubble parameter is constrained, from cosmological data, to be $H_0=67.4\pm0.5$ km s$^{-1}$ Mpc. 

A key incentive to consider modifications of gravity is the evidence for accelerated expansion. While $\Lambda$CDM is a remarkable fit to the data, it is instructive to consider constraints on the equations of state of the form $w_{\rm DE}(a)=w_0+w_a(1-a)$. The most complete analysis using a comprehensive array of data sets \citep{Aghanim2018} find that for a constant equation of state (i.e. $w_a=0$), $w_0=-1.028\pm0.032$ if one uses the Planck CMB data combined with distance measures from supernovae and the BAO in the matter power spectrum. If one allows for $w_a=0$, the same collection of data sets lead to $w_0=-0.961\pm0.077$ and $w_a=-0.28^{+0.31}_{-0.27}$ while using the Planck data, 
the BAO and redshift space distortion measurements of growth from galaxy surveys and galaxy weak lensing data lead to $w_0=-0.76\pm0.2$ and $w_a=-0.72^{+0.62}_{-0.54}$. There is clearly some lee-way in terms of an evolving equation of state but, as discussed above, this can't necessarily be used to find any evidence for modifications to General Relativity.

\subsection{Minor anomalies.}
Given these remarkable constraints and the success of $\Lambda$CDM, it is useful to look at the (minor) inconsistencies and anomalies. While I do not yet believe they are at the level that call into question the success of the standard model, they will be useful in interpreting constraints on gravity later on in this section. There have been a number of small internal anomalies flagged in the current CMB data: large scale surpression of power and evidence for mild anisotropy in the low quadropoles \citep{Schwarz2015}, mild inconsistency between large angle and small angle angular power spectrum constraints of cosmological parameters \citep{Spergel2015}, a discrepancy in the constraint on $\sigma_8$ from primordial and cluster abundances from Sunyaev-Zeldovich measurements \citep{Ade2014}, etc. 

A persistent anomaly is the discrepancy in the constraint on $H_0$ from local measurements, such as those using Cepheids in \citep{Riess2016} as compared to cosmological measurements from, for example, the CMB \citep{Aghanim2018}. While both groups claim percent level constraints on the expansion rate, the discrepancy between the two results is much greater, at a level of reasonable significance. Most attempts at coming up with an explanation for this discrepancy from exotic physics do not explore deviations from GR and I will therefore not pursue this anomaly here.

The only anomaly I would highlight here is the inconsistency between the amplitude of lensed CMB fluctuations as inferred directly from the angular power spectrum of temperature fluctuations and the power spectrum of the lensing potential  reconstructed from the maps \citep{Aghanim2018} . In particular, if ones free up the amplitude, $A_L$ (such that $A_L=1$ is the $\Lambda$CDM value) of the lensing contribution that smooths the the angular power spectrum of temperature fluctuations (as described in Section \ref{cosmology}), current constraints from Planck give $A_L=1.243\pm0.095$ at the $68\%$ confidence level. Paradoxically, if one attempts to reconstruct the power spectrum of the lensing power spectrum, $C^{\phi\phi}_\ell$, directly from the maps one finds that it is consistent with what one would expect from $\Lambda$CDM; i.e. two different methods for estimating the lensed CMB from the same data set give different results.

As mentioned in Section \ref{modgrav}, a key observable to test gravity is the growth rate of structure, i.e. the speed with which gravitational collapse ensues. The main method for measuring the growth rate is with Redshift Space Distortions (RSD). RSDs can be seen in the redshift space correlation function (or power spectrum): along the line of sight, the proxy for distance -- the redshift -- will have a correction from peculiar velocities due to the local gravitational potentials \citep{Kaiser1987}. This means that correlations along the line of sight will be deformed relative to correlations tangential to the line of sight, i.e. the correlation function (or power spectrum) will be anisotropic. On linear scales -- scale of tens of Megaparsecs -- the correlation function, or power spectrum is squashed while on smaller, non-linear scales, they are stretched.

A mild (and arguable) anomaly is the evidence for low growth rate coming from RSDs. First highlighted in \citep{Macaulay2013} but also thoroughly refuted in \citep{Samushia2014}, the discrepancy can best be assessed in Figure \ref{growth} where we can see the {\it predicted}, density weighted growth rate for $\Lambda$CDM, marginalized over a host of cosmological parameters and constrained by the Planck 2018 data \citep{Aghanim2018}. Two things are of note: first, that current low redshift constraints (from RSDs) are weak compared to predicted CMB constraints and second, that many data points seem to be below the predicted line. I emphasize that the evidence for the low growth rate is not yet significant and, for some, not even indicative. 

\begin{figure*}[h]
\includegraphics[width=3in]{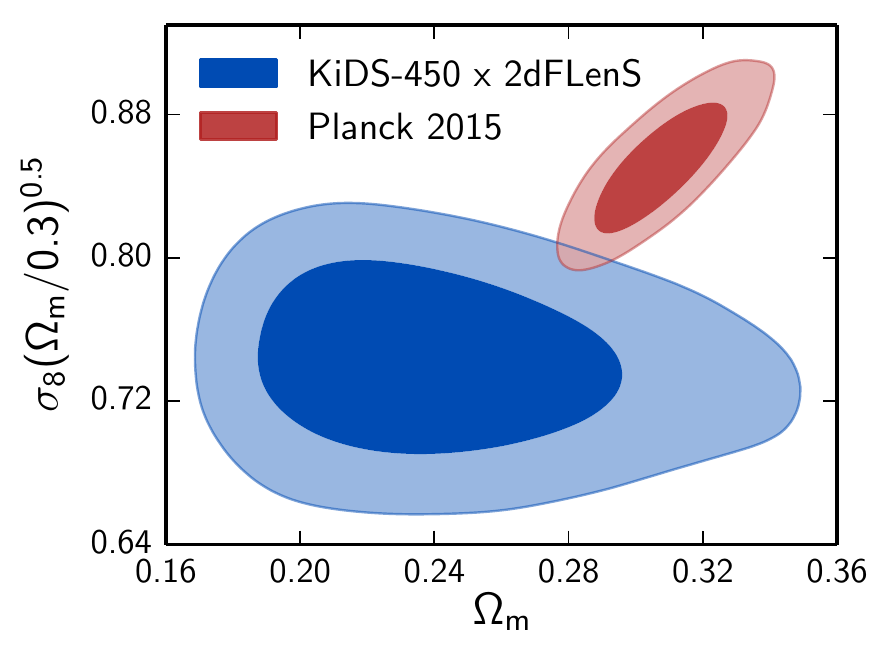}
\caption{Constraint on $S_8$ from the Planck 2015 data (red contours) and the KIDS-450 data (blue-contours) from \citet{Joudaki2018}.}
\label{KIDS_S8}
\end{figure*}

A much more significant inconsistency arises in the comparison between galaxy weak lensing constraints and the CMB. If we focus on $S_8=\sigma_8(\Omega/0.3)^{1/2}$, we can see in Figure \ref{KIDS_S8}, that the galaxy weak lensing measurements from the KIDS survey \citep{Joudaki2018} leads to a somewhat lower value of $S_8$ than inferred from the Planck CMB measurements. This discrepancy is present in the DES data \citep{Abbott2018} although at a milder level; it is possible to further soften the difference with a more elaborate likelihood which, one hopes, can capture the statistical and systematic uncertainties more accurately. But, crucially, these two, completely independent, imaging surveys  both find this inconsistency and combining them, or including more data  may only reinforce this discrepancy.

So, in summary, current data consistently points to $\Lambda$CDM, which is firmly rooted in GR. There are some mild inconsistencies which are not statistically very significant but might lead one to explore deviations from the standard model. We can now look at what current data has to say about extension of GR. 

\subsection{Parameterizing gravity.}
In Section \ref{modgrav} we discussed how to parametrize deviations from GR, in the linear regime, on large scales. In the case of scalar-tensor theories described by the Horndeski action, this is straightforward to do in terms of the five $\alpha_X$ parameters on both super-horizon and sub-horizon scales. It is also true that most of the statistical power will come from smaller scales where, generally, the quasi-static approximation can be used. This greatly simplifies the analysis as one is restricted to two free function, $\mu$ and $\gamma$. In most of what follows, we will present constraints in terms of these parameters or combinations of them. But, before we do, a few comments are in order.

Most parameter constraints using $\mu$ and $\gamma$ assume time dependent, but scale free, functions which are valid from sub-horizon to super-horizon scales. This is clearly not correct, in the sense that it does not naturally emerge from the approach described in Section \ref{modgrav}: one should, at least, expect a scale dependence at or greater than the horizon scale but more generally, the quasi-static approximation breaks down and one needs to, correctly, include the dynamics of the extra degree of freedom (which will, for example, depend on its initial conditions). Having said that, most constraints from LSS are, effectively, on sub-horizon scales (where there is more statistical power) and so one hopes that any inconsistencies in modelling very large scale deviations will be vastly supressed by the data on smaller scales. Effects on the scale of the horizon may be important for large enough surveys (although see \citet{Baker2015} for how insignificant the constraining power of horizon scale modes is) or in the case of the primordial CMB (although these occur at $z\simeq 1000$ and thus are unaffected by late time deviations from GR). The advantages of focussing on $\mu$, $\gamma$ outweigh the problems just described. Nevertheless, later on we will look at constraints on scalar-tensor theories where the large scale behaviour is correctly taken into account.

Another, important, point needs to be highlighted. If we restrict ourselves to scale free  $\mu$ and $\gamma$, one still needs to choose an appropriate time dependence. Strictly speaking, these function will depend on $a$, ${\dot a}$ and, in the case of scalar-tensor theories, the background scalar field value, $\varphi_0$, all of which depend on $t$. A common approach has been to assume that deviations from GR are tied to the onset of accelerated expansion which can be encapsulated in the effective value of the background energy density, $\Omega_{\rm DE}$ or $\Omega_U$ \citep{Ferreira2010} in the notation introduced in Section \ref{modgrav}. So, for example, one can take $\mu=1+\alpha\times(\Omega_{\rm DE}/\Omega_{\rm DE0})$ where $\Omega_{\rm DE0}$ is the value of $\Omega_{\rm DE}$ today, $\alpha$ is a constant and one often refers to $\mu_0\equiv\mu(z=0)=1+\alpha$ (and likewise for $\gamma$). 
A more general approach is to divide the temporal dependence of $\mu$ and $\gamma$ up into uncorrelated redshift bins -- for example, there will be a low redshift bin starting off at $z=0$ followed by a succession of bins. In fact, this approach is often generalized to include scale dependence, i.e. bins in the fourier mode, $k$, as well as $z$.

There are a few drawbacks to any choice of time dependence. First of all, it has been argued that the form $\mu=1+\alpha(\Omega_{\rm DE}/\Omega_{\rm DE0})$  (or of any parameters dependent on $\Omega_{\rm DE}$) does not necessarily reflect the evolution of $\mu$ and $\gamma$ one finds in specific models \citep{Linder2016}. While this may be true in some specific cases,  this time dependence is, generally, a rough approximation of what one might find in theories in which the accelerated expansion is tied to deviations from GR. On the other hand, dividing up the temporal in redshift bins often leads an ill-posed problem (i.e. to many parameters to fit, given the data) and for which results are difficult to determine. Both of theses approximations can be sharpened if, for example, one considers a more systematic expansion of $\mu$ and $\gamma$ in terms of some basis functions, $f_{n}(\Omega_{\rm DE})$ for example, that can capture more elaborate time dependence (for example choosing $f_n(x)=x^n$ or $f_n(x)=j_n(x)$ where the $j_n$ are spherical Bessell functions) or by considering some form of Principal Component Analysis of the time dependence of these functions.  One then has to be careful in choosing priors that adequately reflect the physical behaviour of this functions \citep{Crittenden2012,Espejo2018}

A second drawback then emerges in that the results one finds are, for current data, strongly dependent on the parametrization one considers (or for example the number of free functions one incorporates into the parametrization). For example, with current data it is possible to constrain a constant $\mu$ at the percent level while $\mu=1+\alpha(\Omega_{\rm DE}/\Omega_{\rm DE0})$ is constrained, at best, at ten to twenty percent level (an example of this will be seen later on, when we focus on the BOSS data). Having declared the problems, we will stick with the conventional parametrization in the discussion that follows. 

Note that most current analysis combine multiple data sets -- e.g. CMB with galaxy weak lensing -- so the division between the different results that follow us somewhat artificial. Nevertheless I choose to do so to highlight the impact of the different observables, and their specific effects, on constraints on gravity.

\begin{figure*}[h]
\includegraphics[width=0.9\textwidth]{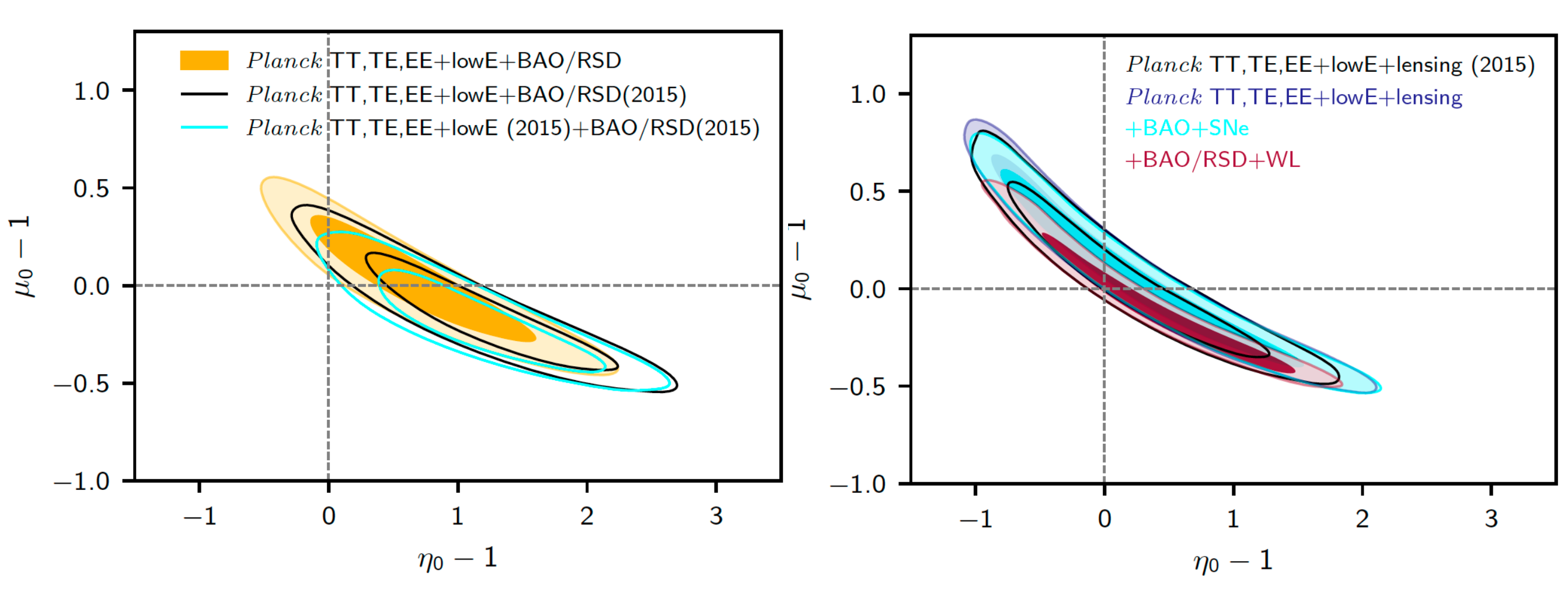}
\caption{Plot from \citep{Aghanim2018}. Constraints on $\mu_0$  and $\eta_0$ (${\tilde \mu}$ and $\gamma$ in this review). On the left hand plot we can see the effect of the $A_L$ anomaly leading to a discrepancy with GR. On the right hand plot, with the inclusion of lensing estimated from the maps, and galaxy weak lensing from DES, consistency with GR is restored.}
\label{planck_constraints}
\end{figure*}

\subsection{Current constraints: the CMB}
If I take a selection of constraints on deviations from GR, I naturally begin with the analysis of the Planck data (combined with other data sets). In \citep{Ade2016} and in \citep{Aghanim2018}, a comprehensive analysis of different parametrizations was undertaken with the 2015 and 2018 Planck data releases; the overall take home message was that current cosmological data is broadly consistent with GR. But of particular interest (and representative of the analysis as a whole) are the features seen in Figure \ref{planck_constraints} from \cite{Aghanim2018}, where constraints on ${\tilde \mu}$ (here labelled $\mu_0$) and $\eta$ are presented.  To begin with we can see that there is a degenerate direction along which constraints are much broader - this direction is set by constraints on the $\Phi+\Psi$ from CMB lensing and thus on $\Sigma$ (if one transforms to the $\mu$, $\Sigma$ parametrization, one can, to some extent decorrelate the two parameters being constrained). 

Second, we note that the error bars on these parameters are of order $1$. While these may seem very weak compared to the precision one obtains on astrophysical scale (for example, the constraints on the analogous parameter to $\gamma$ from the Solar System is of order $10^{-5}$) they are obtained in a completely different regime, on scales which are fifteen orders of magnitude greater and where the curvature scale is  twenty orders of magnitude smaller (see Figure \ref{BPS}).

Finally, and most striking, is that fact that there seems to be a  detection of deviations from GR, away from $\mu_0=\gamma_0=1$. This push away from GR from the Planck data alone is due to the enhanced amplitude of CMB lensing anomaly (the "$A_L$ anomaly") previously discussed in this section, as can be seen in the left hand panel of Figure \ref{planck_constraints}. In the right hand panel we see that,  if one then includes the measurement of the lensing potentials from the maps -- $C^{\phi\phi}_\ell$ -- (which, recall from the discussion above, is an independent measurement to the one used to determine $A_L$), the constraints are pushed towards GR. Including galaxy weak lensing (in this case from DES) which, as we saw, favours a lower amplitude in the lensing potentials, further pulls the constraints towards GR. Adding the RSD constraints has a small effect in the orthogonal direction.

Is this a significant detection of deviations from GR? Not if one properly accounts for the increase in complexity of the model, even though there is an improvement in the goodness of fit.
GR would have to be excluded at a much higher significance for the detection to be noteworthy. 

\begin{figure}
\centering
\includegraphics[width=.9\textwidth]{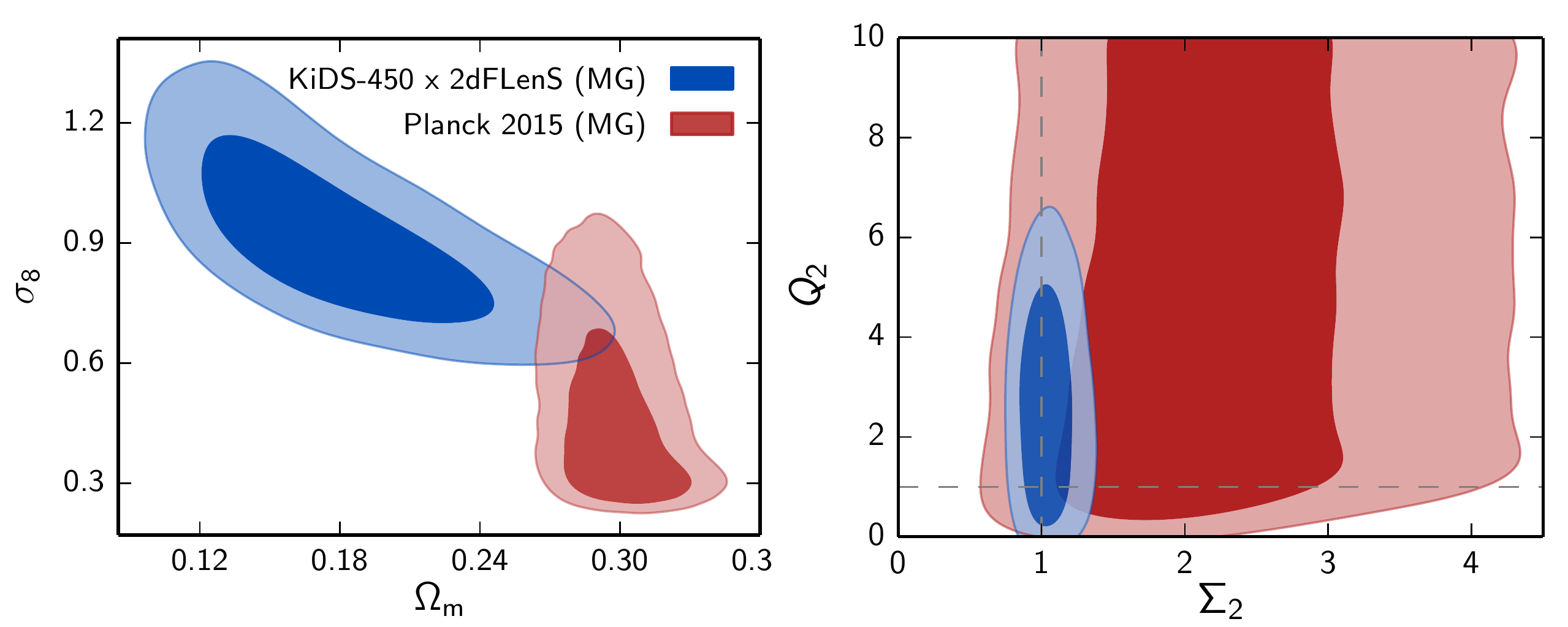}
\caption{ Constraints from the KIDS-450 data combined with RSD measurements from 2dFLenS and BOSS (in the overlap region with KIDS-450) and other data sets. The right hand plot shows constraints on $Q_2$ (defined as ${\tilde \mu}$ this review)  and $\Sigma$. The left hand plot shows constraints on $\Omega_{\rm m}$ and $\sigma_8$; adding ${\tilde \mu}$ and ${\Sigma}$ alleviates the discrepancy in $S_8$.}
\label{kids_constraints}
\end{figure}

\subsection{Current constraints: galaxy weak lensing and growth rate.}

We now turn to a different set of constraints, obtained from a combination of galaxy weak lensing from the KIDS-450 with the CMB and RSDs \citep{Joudaki2018} . In this case, the parameters being constrained are ${\tilde \mu}$ (here dubbed $Q$) and $\Sigma$, binned in both redshift and scale. If we look at the constraints for the lowest redshift bin (labelled by a "2"), on largest scales (assume that we are marginalizing over the other bins) and focus on the case of CMB alone (the red dashed lines in the right hand panel) we see, again, a mild inconsistency with GR; in this case it is milder than in the Planck analysis and comes tied to vastly inflated uncertainties (as is expected from our discussion above, given the number of free parameters -- or bins -- now included). Weak lensing and RSD data (without CMB) leads to tighter constraints, centered on the GR value. To combine them with CMB data, the authors have introduce a large scale cut to mitigate the effect of non-linear scales. In that case, the constraints including the CMB are completely consistent with GR; again the low lensing potentials from galaxy weak lensing compensate the high $A_L$ constraint from the CMB.
An interesting effect of enlarging the parameter constraints to include deviations from GR is that the "$S8$" discrepancy can be substantially mitigated. In the left hand panel of Figure \ref{kids_constraints} one can see that what were orginally disjunct contours (the dashed lines) can be rendered overlapping (the teal and red shaded contours) by adding in $\mu$ and $\gamma$. We note that the Dark Energy Survey (DES) has undertaken an analysis of their first year of data \cite{Abbott2018DES}  to come up with constraints on $\mu$ and $\Sigma$. Their constraints are of the same order of magnitude as the KIDS-450 constraints although marginally more consistent with the GR values. 

\begin{figure*}[h]
\includegraphics[width=3in]{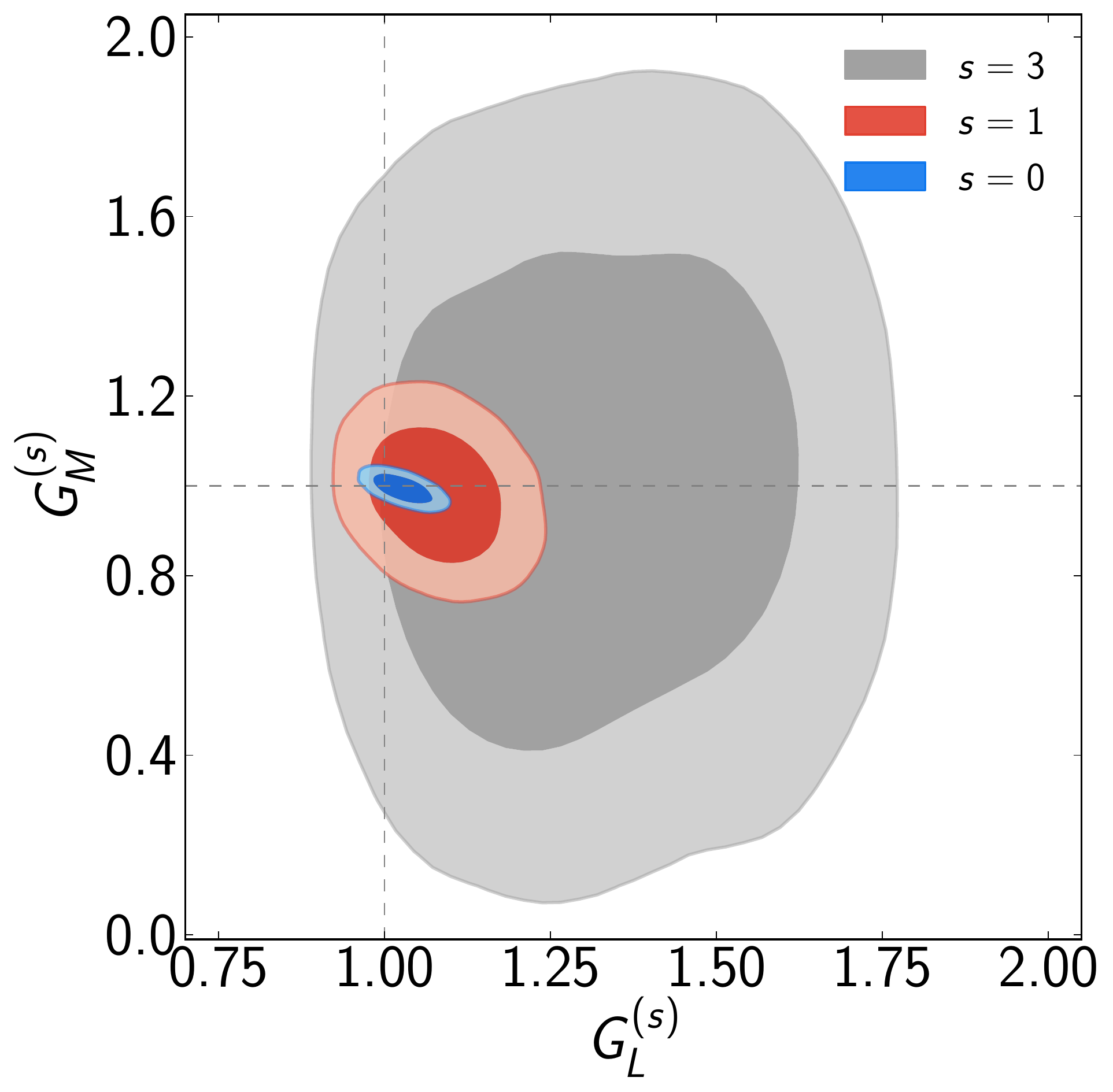}
\caption{Constraints of $G_M$ and $G_L$ using BOSS measurements of the growth rate  combined with other data sets (figure kindly produced by E.\,M.\,Mueller). A parameterization of the form $G_X=1+(G_X^{(s)}-1)a^s$ is assumed with $s=0,1,3$.}
\label{BOSS}
\end{figure*}

We now briefly look at the SDSS-III
BOSS constraints in \citep{Mueller2018} where the authors concentrate on $G_{M}$ and $G_{L}$ which, we know from Section \ref{modgrav}, can be directly related to $\mu$ and $\gamma$. Assuming a time dependence for these quantities of the form $G_X=1+(G_X^{(s)}-1)a^s$ where $X=M,L$ and, in their analysis, $s\in\{0,1,3\}$. In Figure \ref{BOSS} we see how a time independent choice ($s=0$) leads to very strong constraints while $s=1$ (and even more so with $s=3$) leads to weaker constraints as deviations from GR are more suppressed at high redshift. 
 Constraints on the growth index are found to be $\gamma=0.566\pm0.058$ which, again, show strong consistency with GR. 

It is important to note that the analysis of  data to constrain $f\sigma_8$ involves a number of steps in which a particular cosmology and theory of gravity is assumed. A concern, then, is that estimates of $f\sigma_8$ may be biased by these assumptions. Attempts at estimating the size of this bias \citep{Barreira2016,Bose2017} show that it is insignificant for current surveys but may be an issue when their statistical power greatly improves.
 
 \begin{figure*}[h]
\includegraphics[width=3in]{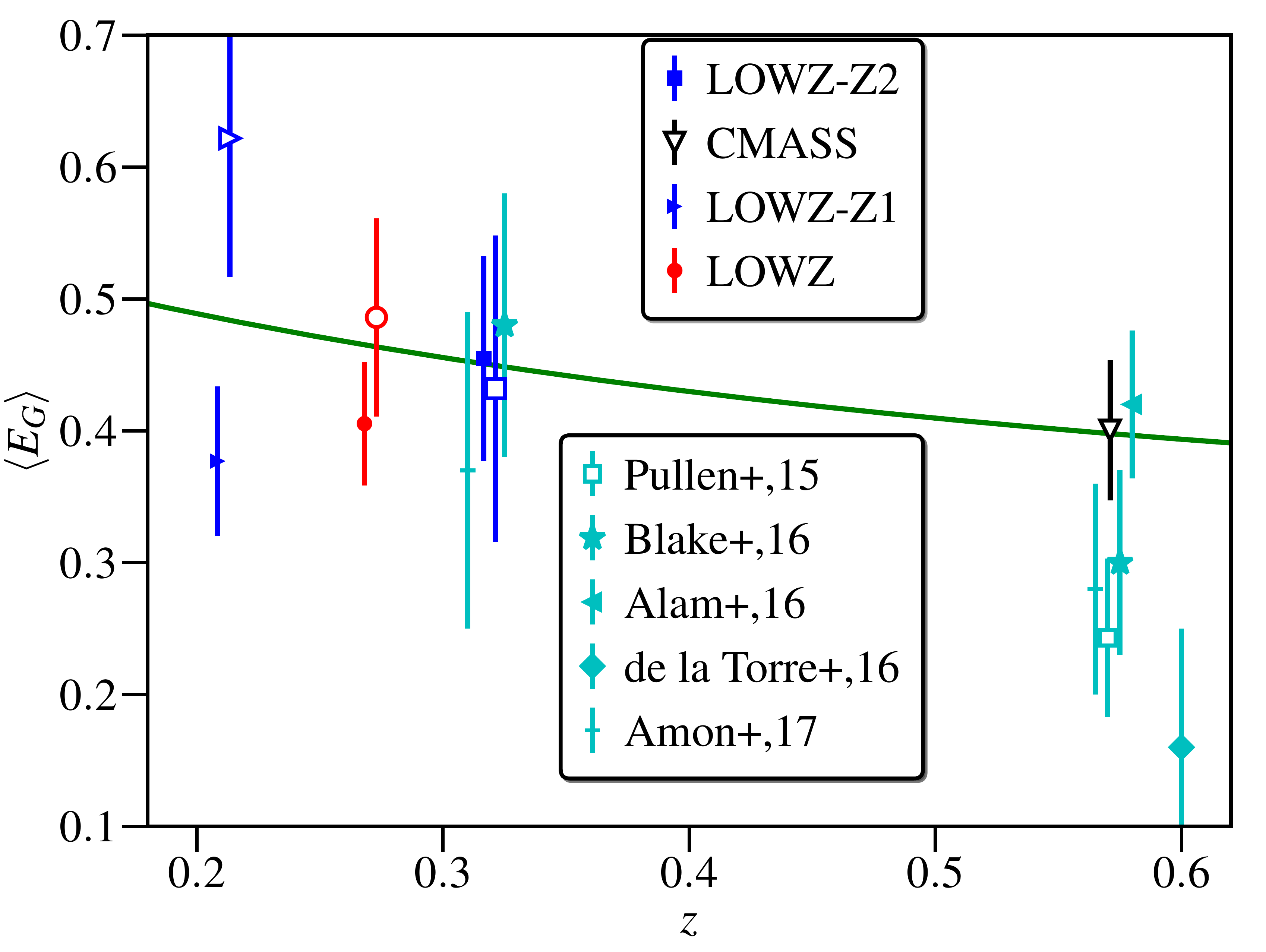}
\caption{State of the art constraints on $E_G$ using both CMB and galaxy weak lensing (figure kindly produce by S. Singh).}
\label{EG_constraints}
\end{figure*}

 A parallel line of attack has been to determine constraints on GR from $E_{\rm G}$. There have been a number of attempts at measuring $E_{\rm G}$ using galaxy-galaxy lensing \citep{Reyes2010,Amon2018} and CMB lensing combined with Galaxy velocities \citep{Pullen2016} finding this statistic to be generally consistent with GR, but with a few minor complications (for a thorough analysis \citep{Singh2018}). First of all, in the case of \citet{Pullen2016}, the large angle contribution of CMB lensing leads to a 2.6$\sigma$ inconsistency with GR. Second, the choice of background cosmology has a strong impact on the estimate of $E_G$; for a comfortable consistency with $\Lambda$CDM, one needs a slightly lower $\Omega_M$ than normally assumed.  The current, state of the art, constraints on $E_G$ are given in Figure \ref{EG_constraints}.
 
It is not clear whether $E_G$ is a particularly powerful statistic for testing GR and whether it adds anything more than the other, more standard likelihood analysis currently being undertaken. While, in principle, it should mitigate the contaminating effect of galaxy bias, in practice and due to the way it is implemented, one is still at the mercy of one's best bias model. Furthermore, a proper, full likelihood analysis of all cosmological data will necessarily include cross-correlations between lensing and density measurements and will inevitably work towards breaking any degeneracy between nuisance parameters such as bias and the GR parameters one is attempting to constrain.  
 
\subsection{Model specific constraints: scalar-tensor theories.}
Instead of using a general, parametrized approach to constraining GR, it is sometimes useful to focus on specific models. This makes it possible to, to some extent, span different regimes and compare, for example, the constraining power of Solar System probes versus cosmological probes. The workhorse for such constraints is Jordan-Brans-Dicke gravity, the simplest scalar-tensor theory described in Section \ref{modgrav} with its one free, dimensionless parameter, $\omega_{\rm BD}$. Current astrophysical constraints of $\omega_{\rm BD}$ are remarkable. From the Cassini time delay it is possible to constrain $\omega_{\rm BD}>40,000$ \citep{Bertotti2003} while from milli-second pulsars the constraints are around $\omega_{\rm BD}>20,000$ \citep{Freire2012}. Current cosmological constraints, using the Planck data are much weaker: $\omega_{\rm BD}>900$ and come primarily from the effect of the extra scalar field on the background expansion and thus on the angular diameter distance \citep{Avilez2014}. Measurements of the growth rate are not yet up to the level of precision required to lead to competitive constraints on $\omega_{\rm BD}$.

Jordan-Brans-Dicke gravity is not screened on astrophysical scales but more general scalar-tensor theories will be. For example Galileon theories may have Vainshtain screening while conformally coupled theories with potentials may have Chameleon or symmetron screening. If that is the case, then the {\it only} way of obtaining tight constraints may be in the cosmological regime (although see Sections \ref{screening}  for other approaches). It would make sense then to look at constraints on a general scalar tensor theory such as the one described by the Horndeski action. Furthermore, as we saw above, the $\mu$, $\gamma$ parametrization shouldn't be extended to horizon and superhorizon scales (although these scales won't have a strong statistical weight and will contribute little to the final results). The Horndeski action allows a completely self consistent parametrization on all scales in terms of the $\alpha$ parameters.

The most comprehensive analysis of the Horndeski action was undertaken in \citep{Bellini2016}, where the authors used CMB data (a combination of the 2013 Planck power spectra and WMAP polarization), galaxy redshift survey data in the form of the power spectrum (from WiggleZ), the BAO (from BOSS, WiggleZ and 6DFGS) and RSDs (from SDSS, BOSS, WiggleZ, 6DFGS and VIPERS) to place constraints on the $\alpha$s. In their analysis, the authors used a parametrization of the form $\alpha_X=c_X\Omega_{\rm DE}$ (where $c_X$ is a constant), along the lines of the one use for $\mu$, $\gamma$. On the scales being constrained, $\alpha_K$ was found to be irrelevant and could be fixed to fiducial values without affecting the final results. The typical variance on $c_X$ was found to be ${\cal O}(1)$ and the parameters were found to be mostly uncorrelated. 

Interestingly, the analysis of \citep{Bellini2016} found that the best fit values of $\alpha_B$ and $\alpha_T$ were away from GR and, for some combination of data sets, at the $95\%$ confidence level. While for CMB alone, the results were perfectly consistent with GR, the inclusion of RSDs pulled the preferred values in such a way that $c_T$ was preferentially less than zero and $c_B$ was preferentially greater than zero. This is somewhat at odds with the analysis of $\mu$, $\gamma$ in \citet{Aghanim2018} where it is the $A_L$ anomaly that is pushing the constraints away from GR and where the RSDs do not play a significant role. The authors do not, however, claim a significant detection of deviations from GR on the grounds that, having introduced more parameters, the evidence (which they compare to GR using the Bayes factor) is not sufficiently significant. In other words, the increase in complexity in choosing Horndeski is not compensated by the improvement in the fit to the data.

A restricted analysis of Horndeski gravity taking into account the constraints from GW170817 on the speed of gravitational waves has been undertaken in \citet{Kreisch2017} . The authors found that, varying $\alpha_B$, the values of $\alpha_M$ were discrepant with GR at the $95\%$ confidence level (but fully consistent if $\alpha_B=0$)  and they argue that this is due to the fact that, with the specific choice of kineticity $\alpha_k=0.1$, an artificial prior is induced which biases the results.
More recently. the analysis of the Planck 2018 data \citep{Aghanim2018}, for the parameter subspace where $\alpha_B=-\alpha_M$ found slight preference for $\alpha_M<0$ which disappeared once RSD and WL data was included, consistent with their analysis in term of $\mu$, $\gamma$.

Of particular interest are the current constraints on Galileon theories. The constraints on the speed of gravitational waves from GW170817 rules out theories with quartic or quintic terms in the action of equation \ref{eq:L_horndeski}. Constraints from large scale structure and, in particular, the Integrated Sachs Wolfe effect in equation \ref{CMBeq} can be used to show that the cubic terms lead to a grossly discrepant effect \citep{Renk2017}. This means that cosmological Galileon gravity is effectively ruled out.

As we have seen, they current situation is that constraints on different parameterizations of deviations from GR on cosmological scales are still weak. Yet, in particular situations there is mild evidence for non GR behaviour. In one case, this evidence is primarily driven by the high $A_L$ anomaly in the angular power spectrum of the CMB and can be compensated by the low amplitude of galaxy weak lensing or RSDs. In another analysis, in this case of Horndeski theories, it is the RSDs that are pulling the constrained values of the $\alpha$s away from GR. Furthermore, the results depend quite significantly on the time dependence assumed or the cuts to which the data sets have been subjected. Given that the attempts at constraining deviations from GR are so disparate in methodology, as well as few and far between, there is a serious concern that a complete and final set of reliable, calibrated results is still lacking and that can be taken as a basis for more definitive statement about the status of GR on cosmological scales.  Furthermore, the overriding consensus is that there is no evidence for deviations from GR from a Bayesian evidence point of view: the complexity of the extensions outweighs the improvement in the fits.

Nevertheless, these various analysis' have shown the way forward in how to constrain gravity on cosmological scales. A clear approach to the analysis is emerging and, more importantly, a sense of the systematic problems and pitfalls that must be dealt with if one is to obtain significant constraints of GR with current and future data.

\section{Alternative constraints - clusters, galaxies and voids.}
\label{alternatives}
As we have seen, there has been some progress in applying the techniques for constraining GR on large scales to existing data, with mixed results. The situation is less developed on smaller scales where, on the one hand, non-linear gravitational collapse and baryonic physics can play a significant role and, on the other hand, the distinctive signatures of screening can emerge. Existing constraints are fewer, more scattered and, unlike on large scales, very model specific. In this section we will undertake a brief survey of the more durable results.

\subsection{Cluster masses.}
To begin with, and as we saw discussed in Sections \ref{modgrav} and \ref{screening}, it makes sense to look for evidence for gravitational slip by comparing the way relativistic particles (i.e. photons) and non-relativistic particles (e.g. baryons or dark matter) respond to a gravitational potential and mass. One such approach is to compare different ways of measuring cluster masses: X-ray, Sunyaev-Zeldovich and weak lensing observations. Each one of these observations measures different things, from temperature and pressure profiles to gravitational potentials, and it is necessary to construct a model (using, for example the fluid equations and assuming hydrostatic equilibrium) to bring them all together. Such an approach has been undertaken in \citet{Terukina2014} and used to place constraints on the different free parameters in an individual cluster, including one that controls deviations from gravity. The calculation is done in the non-linear regime and has been done for one particular model, $f(R)$; the authors found $df/dR \lesssim 5\times 10^{-5}$. This method has also been applied to constraints on the Vainshtein mechanism where no evidence for screening was found \citep{Sakstein2016}.

Instead of looking at individual clusters, one can look at the statistics of ensembles of clusters and, for example, the cluster mass function, $n(M)$. Arguably, this statistic will not be overwhelmingly contaminated by baryonic physics given the size of the objects and the scales probed. Attempts at constraining this statistic necessarily involve N-body simulations (although progress has been made developing semi-analytic methods) and are, again, model specific. Again, most effort has gone into testing $f(R)$ (and some other alternatives) using  estimates of cluster abundances from a range of cluster catalogues \citep{Schmidt2011,Lombriser2012,Cataneo2015}. The constraints are at the level of $df/dR \lesssim 1-10\times 10^{-5}$. 
\subsection{Screened fifth forces in galaxies.}
Some progress has been made in looking for screening within galaxies. The latest, most detailed search for chameleon screening  has, arguably, lead to evidence of a screened fifth force  of the form $F=-\Delta G e^{-mr}/r^2$ with a strength of $\Delta G/G\simeq 0.01$ on scales of $1/m\sim 1$ Mpc. This evidence comes from two very different types of analysis. In the first case, a systematic offset between the screened stellar component and the unscreened HI component of a selection of galaxies is consistent with the presence of a weak fifth force \citep{Desmond2018a,Desmond2018b}. A number of systematic effects have been included and are taken into account in the error budget although two caveats most be considered. If one considers a (overly) conservative error on the centre of the HI gas (effectively doubling the uncertainty that is presented in the data  set), the detection goes away and one obtains a strong upper bound,  $\Delta G/G<0.01-0.1$ depending on the screening scale. But, and more importantly, little is know about how baryonic effects may lead to a signal that mimics the fifth force. While the signature that is constrained is very specific, a detailed analysis of ultra-high resolution simulations of galaxies in a large enough volume is required to completely confirm or rule out the detection.

The second type of the analysis looks at another signature described in Section \ref{screening}: warped stellar disks as the galaxies fall into their (displaced) dark matter halos \citep{Desmond2018d}. There the authors used r-band images of mostly late time galaxies in the NASA Sloan Atlas to compile a catalogue of approximately $10^3$ warps with a specific focus on the "cup" shaped signature of a screened fifth force. Again, they found a strong indication of a fifth force with $\Delta G/G \simeq 0.01$ on scales of around $2$ Mpc, surprisingly consistent with the measurement of star-gas offsets. Remarkably, this new analysis is with an altogether different data set and involving a very different signature. Nevertheless it is potentially plagued by its own, as yet, unquantified systematics: the effects of baryonic physics is poorly understood but more importantly, there is no clear idea yet of how warps arise in galaxy. While there is a consensus that warps are induced by environmental effects, the fact that warps have been found in isolated galaxies calls this environmental origin into question. A better understanding of how warps form would greatly help in firming up this detection of a screened fifth force.

A different, but related, constraint on screened forces has been found by attempting an Eotvos like test on galactic/cosmological scales. In essence, the idea is to compare the motion of a "small" stellar system and a supermassive black hole in the gravitational potential of another, larger system. A black hole is chosen as one of the test bodies because in some theories \citep{Hui2012}, it will satisfy a no-hair theorem and therefore will be unscreened (the authors are inspired by, and very specifically address, the case of Galileon theories which are endowed with Vainshtein screening).  As a result, and for a reasonable choice of the screening parameters, one should expect an offset of ${\cal O}({\rm kpc})$ between a black hole and its host galaxy.

In \citep{Asvathaman2017} and \citep{Sakstein2017}, the authors looked at the black hole centred on M87 as it falls under the influence of a sub-clump of the Virgo cluster around M84 and M86. They found that the fifth force field strength of the Galileon must be less than $10^3 (km/s)^{2}/kpc$. The authors had to make a number of assumptions about the mass to light ratio, concentration, halo profile, etc, but ultimately found that their constraint is stronger than current constraints on Galileons from the Solar System. Again, much is to be done to better understand the systematics and it is clear that more robust constraints will be obtained with more objects but the route is promising for constraining (or detecting) Vainshtein screening.
\subsection{Cosmic voids.}
The arena, par excellence, where constraints on the screening mechanism should be most effective, is in cosmic voids. As explained in section \ref{screening}, one should expect forces to be unscreened as they are regions of low density, curvature or potential. There has been a reasonable amount of effort in thinking about what signatures to look for, primarily based on analytic calculations and numerical simulations. The use of voids as a tool of more standard cosmological analysis is also growing with first attempts at measuring the void-correlation function already in hand. 

Surprisingly, there are no {\it de facto} constraints on deviations from GR with current data. The only attempt, thus far, is embedded in efforts to constrain cosmological parameters from the void correlations function. The authors of \citep{Hamaus2016} have found a constraint on the linear $f/b=0.417\pm0.089$ at median redshift, ${\bar z}=0.57$, which is entirely consistent with $\Lambda CDM$. Thus, they claim, there is no evidence for deviations from GR within voids.

\section{Future prospects. }
\label{future}
\subsection{Theoretical status.}
Testing gravity with cosmology has come a long way over the last decade. And, in some aspects, one might consider it a mature field. For a start, we have a much clearer understanding of what it means to construct alternatives to General Relativity. Primarily it involved adding extra degrees of freedom but do so one has to be careful that instabilities, ghosts or other pathologies aren't unwittingly introduced. And even if they are, they may turn out to be benign. So, while in the past, there were a set prejudices and a disparate set of rules of what one could or could not do when tampering with GR, there is now a much more coherent and sensible approach. One of the by products is that we have a clearer understanding of the landscape of gravitational theories and how they link together. For example, scalar-tensor theories take pride of place and are consumately well understood even though that understanding is not necessarily complete. And their structure can be mimicked in vector-tensor theories leading to a burgeoning exploration of this different part of the landscape. Tensor-tensor theories -- massive gravity or bigravity theories -- will reduce, in a particular sector to a type of scalar tensor theories but also have properties that are uniquely their own. And so on as one moves into higher dimensions and non-locality.

From the point of view of cosmology, and in particular on the largest scales, there is now a completely well understood theory of a linearly perturbed expanding universe. So much so that it can be (and has been) implemented in precise Einstein-Boltzman solvers that allow us to predict (linear) cosmological observables with exquisite precision. To a lesser extent, we also have a good idea of how smaller scales may be subjected to gravitational screening mechanisms that can mask fifth forces; we have a reasonable complete view of the different possible mechanisms and have numerical implementation of some of them. 

\subsection{Observational status.}
When it comes to obtaining constraints from current data, the situation is more uncertain. The first steps have been taken and there are now a number of pipelines that can take current observations of the large scale structure of the universe -- the CMB, galaxy and weak lensing surveys -- and find a set of constraints on gravitational parameters, i.e. parameters that encapsulate deviations from GR on linear scales. The results are intriguing in that anomalies in the CMB power spectrum due to what seems like excess gravitational lensing push the constraints {\it away} from GR. Paradoxically, the low amplitude of lensing in galaxy gravitational lensing surveys can compensate when combined with the CMB data. Measurements of growth seem marginally lower than what is expected from GR yet do not yet have enough statistical power to play significant role.

An overall concern is the disparate set of methods and parametrizations being used by the (few) different groups which makes it difficult to compare and, more importantly, cross calibrate the different results. This differs from cosmological constraints of parameters {\it within} $\Lambda$CDM; there, a large number of groups have found constraints on, for example, the fractional densities of the different energy densities, the expansion rate, the spectral index of fluctuations, etc and there is a very clear and accurate consensus of what they are. Until this level of rigor is applied to the gravitational parameters, it is difficult to interpret what the current constraints actually mean or whether they are robust.

Nevertheless, where there are constraints, they are still weak, roughly of ${\cal O}(1)$. This means that, even in the case where there is a 3$\sigma$ discrepancy with GR, the increase in complexity by including extra gravitational parameters still outweighs the improvement in the likelihood. This is typically characterized in terms of Bayes factors, comparing the Bayesian evidence between the two different scenarios. The fact that, from this point of view, deviations from GR are disfavoured relative to GR parallels other attempts at extending $\Lambda$CDM and merely confirms that the base model is still the best candidate for explaining the Universe.

On smaller scales, deeply into the non-linear regime where more complex, baryonic physics can play a role, tests of gravity are still in their infancy \citep{Cataneo2018}. While it is an extraordinarily difficult regime to work in and prone to countless systematics, it is also, potentially immensely rewarding. Data on these scales is accruing at a phenomenal rate and, if systematics are properly modelled and under control, they should allow for constraints (or detections) with a high statistical power. The first steps have been taken with a few detections and constraints of an altogether new type of effect -- gravitational screening -- which have emerged. There is clearly a vast and uncharted territory to be explored.

\subsection{Planned surveys and prospects.}

Although current constraints are weak and inconsistent, the hope is that with the new generation of surveys, the situation will change dramatically. Quite possibly, over the next decade, we will transition from ${\cal O}(1)$ constraints on deviations from GR to ${\cal O}(10^{-1})$ or even ${\cal O}(10^{-2})$. A number of planned experiments dominate the landscape. The Euclid satellite ({\tt http://sci.esa.int/euclid/}), primarily funded by the European Space Agency, will survey 15,000 square degrees of the sky, building up a catalogue with spectra of upto $10^7$ galaxies. In parallel, a ground-based spectroscopic survey, DESI ({\tt http://desi.lbl.gov}), based in Arizona will collect spectra of galaxies of a similar number and area. The Large Synoptic Survey Telescope, LSST,  ({\tt http://www.lsst.org}) based in the Atacama Desert will undertake a deep, imaging, survey ($r>27$)  with an area of up to 20,000 square degrees. The Square Kilometre Array, SKA, ({\tt  https://www.skatelescope.org}) will produce a radio survey of HI emission and galaxies in the frequency range $50-1760$ MHz, mapping structure out to $z\simeq 3-4$. The Simons observatory ({\tt https://simonsobservatory.org/index.php}), also in the Atacama Desert will map the CMB to exceptionally high resolution and sensitivity, while the Stage IV CMB survey ({\tt https://cmb-s4.org}) will, hopefully, be a coordinated set of multiple ground based observatories which will cover upto $40\%$ of the sky.

One can forecast the precision with which one will be able to constrain the gravitational parameters. For example, with Euclid or LSST, we expect to reduce the uncertainty in $\mu$ and $\gamma$ from ${\cal O}(1)$ today down to ${\cal O}(10^{-2})$ . If we restrict ourselves to scalar-tensor theories described by the Horndeski action, again, and consider combinations of LSST, SKA and Stage IV experiments, we find the uncertainty in the $\alpha_X$ is reduced to ${\cal O}(10^{-1})$ to ${\cal O}(10^{-2})$ \citep{Alonso2017}. An interesting case is that of JBD gravity where we found that current constraints place $\omega_{\rm BD}>10^3$; future constraints, using combinations of LSST, SKA and Stage IV will push $\omega_{\rm BD}>{\rm few} \ \times \ 10^4$, comparable with constraints on the Solar System and astrophysical scales. Once we reach this level, we will truly have entered a new age in gravitational physics, with constraints on cosmological scales playing a significant role in understanding gravity beyond GR. In fact, if the values of $\mu$ and $\gamma$ (or equivalently the current values of the $\alpha_X$) which current data seem to find discrepant with GR hold up, then there will be uncontrovertible evidence of new physics.

Future surveys will also allow us to greatly improve searches for screening, or fifth forces more generally, in galaxies and other astrophysical structures. For example, with the SKA it will be possible to obtain constrains on $\Delta G/G\simeq 10^{-9}$ using HI-optical offsets of galaxies centres \citep{Desmond2018a,Desmond2018b}.

\subsection{Synergies with other gravitational probes.}
With the parallel development of observational cosmology, gravitational wave physics and black hole imaging, interesting (and inevitable) synergies have developed. For a start, and as more and better detections of compact object mergers are recorded, it will become possible to undertake cosmological constraints with such events. A first attempt with GW170817 at constraining the Hubble constant has shown the tremendous effectiveness of this new window on the expansion rate of the Universe \citep{Ligo2017}. The constraint on the speed of gravitational waves has already had a devastating impact on the fauna of modified gravity theories, leading to a severe cull of acceptable models \citep{Baker2017,Creminelli2017,Ezquiaga2017}. Furthermore, the absence (or presence) of friction in the propagation of gravitational waves can have an impact on what deviations from GR are allowed \citep{Belgacem2018,Amendola2018}. And the promise of detecting a stochastic background of gravitational waves is leading to renewed scrutiny of novel effects that might arise in deviations from GR.

A particular example of synergy between cosmological and gravitational wave physics has recently been invoked in \citet{Tattersall2018} where it was shown that many (primarily scalar-tensor) theories in which the extra fields had a cosmological impact, yet in which the speed of gravitational waves was equivalent to that of light, would obey no-hair theorems to some degree. This means that, for example, any smoking gun for deviations of gravity would have to be on large, cosmological, scales or in the, still dynamical, regime post-merger of binary collision events. I.e. these theories would not manifest themselves during the inspiral or in black hole imaging.

This emerging and exciting synergy is the hallmark of what I believe is a new and exciting phase in gravitational physics, a new "golden age" for general relativity. Until now the focus, and quest, for new physics has been in the high energy regime, of particle accelerators and direct (and indirect) dark matter searches. With the multiple new windows on the gravitational universe, in which cosmology takes pride of place, one would hope that new forces and phenomena are on the verge of discovery. In the very least, we will end up with a cast iron theory for gravity, tested over an enviable range of scales and regimes.

\section*{DISCLOSURE STATEMENT}
The author is not aware of any affiliations, memberships, funding, or financial holdings that
might be perceived as affecting the objectivity of this review. 

\section*{ACKNOWLEDGMENTS}
Much of my understanding of this field was developed with a wide number of collaborators, to all of whom I am grateful. I would like to thank specific input from Alex Barreira, Harry Desmond, Shahab Joudaki, Kazuya Koyama, Martin Kunz, Macarena Lagos, Danielle Leonard, Antony Lewis and Alessandra Silvestri. I am also extremely grateful to Eva-Maria Mueller, Shahab Joudaki and Sukhdeep Singh for producing "bespoke" images for this review. This work was supported by ERC Grant No: 693024, STFC and the Beecroft Trust.


\end{document}